\newcommand{\bra}[1]{\ensuremath{\left\langle {#1} \right|}}
\newcommand{\ket}[1]{\ensuremath{\left|  #1 \right\rangle}}
\newcommand{\expec}[1]{\ensuremath{\left\langle {#1} \right\rangle}}

\newcommand{\wpk}{\ensuremath{W_{\mathrm{pk}}}}
\newcommand{\Goz}{\ensuremath{W}}
\newcommand{\Gto}{\ensuremath{\Gamma_{3}\,}}
\newcommand{\TA}{\ensuremath{T_{\mathrm{A}}(\omega)}}
\newcommand{\Tph}{\ensuremath{T_\phi(\omega)}}
\newcommand{\Jperp}{\ensuremath{J_{\perp}^2\;}}

\newcommand{\Ne}{\ensuremath{N_{\uparrow}}}
\newcommand{\Ng}{\ensuremath{N_{\downarrow}}}

\newcommand{\ddt}{\ensuremath{\frac{\mathrm{d}}{\mathrm{d}t}}}

\documentclass[aps,prl,amsmath,amssymb,reprint,showpacs]{revtex4-1}
\usepackage{bibunits}
\usepackage{graphicx}
\usepackage{color}
\usepackage{hyperref}

\begin{document}
\begin{bibunit}

\title{Relaxation oscillations, stability, and cavity feedback in a superradiant Raman laser}
\author{Justin G. Bohnet}
\email[To whom all correspondence should be addressed. ]{bohnet@jilau1.colorado.edu}
\author{Zilong Chen}
\author{Joshua M. Weiner}
\author{Kevin C. Cox}
\author{James K. Thompson}
\affiliation{JILA, NIST and Department of Physics, University of Colorado, Boulder, Colorado 80309-0440, USA }



\begin{abstract}
We experimentally study the relaxation oscillations and amplitude stability properties of an optical laser operating deep into the bad-cavity regime using a laser-cooled $^{87}$Rb Raman laser.  By combining measurements of the laser light field with non-demolition measurements of the atomic populations, we infer the response of the the gain medium represented by a collective atomic Bloch vector.  The results are qualitatively explained with a simple model. Measurements and theory are extended to include the effect of intermediate repumping states on the closed-loop stability of the oscillator and the role of cavity-feedback on stabilizing or enhancing relaxation oscillations.  This experimental study of the stability of an optical laser operating deep into the bad-cavity regime will guide future development of superradiant lasers with ultranarrow linewidths.
\end{abstract}

\maketitle

Optical lasers operating deep in the bad-cavity or superradiant regime, in which the cavity linewidth $\kappa$ is much larger than the gain bandwidth $\gamma_\perp$, have attracted recent theoretical\cite{MYC09,CHE09} and experimental \cite{BCW12, BCW12Hybrid, WCB12} interest.  The interest has been partially driven by the possibility of creating spectrally narrow lasers with linewidths $\le 1$ millihertz and dramatically reduced sensitivity to the vibrations that limit state of the art narrow lasers and keep them from operating outside the laboratory environment\cite{JLL11}. These lasers may improve measurements of time\cite{Ludlow28032008}, gravity\cite{CHR10}, and fundamental constants\cite{PhysRevLett.98.070801,PhysRevLett.100.140801} aiding the search for physics beyond the standard model. The cold-atom Raman superradiant laser utilized here operates deep into the bad-cavity regime ($\kappa/ \gamma_\perp \approx 10^3 \gg 1$), making it an important physics test-bed for fundamental and practical explorations of bad-cavity optical lasers. 

In the interest of fundamental science and in light of the potential applications, it is important to understand the impact of external perturbations on lasers operating deep into the bad cavity regime. In this Letter, we present an experimental study of the response to external perturbations of the amplitude, atomic inversion, and atomic polarization of an optical laser operating deep into the bad-cavity regime. In contrast, experiments have extensively studied the amplitude stability properties of good-cavity lasers ($\kappa \ll \gamma_\perp$) (See Ref. \cite{M66} and references therein). Previous experimental work in the extreme bad cavity\cite{BCW12} and crossover regime\cite{PhysRevLett.72.3815} focused on the phase properties of the light and atomic medium. Amplitude oscillations, intensity noise, and chaotic instabilities have been observed in gas lasers operating near the cross-over regime ($\kappa/\gamma_\perp \le 10$)\cite{C78,EEW98,HB85,WAH88}. Relaxation oscillations of the field have been studied deep into the bad cavity regime using masers\cite{shirley:949} in which the radiation wavelength is comparable to the size of the gain medium, unlike in the present optical system. Previous theoretical studies of amplitude stability deep in the bad-cavity regime include studies of relaxation oscillations\cite{PhysRevA.47.1431}, chaotic instabilities\cite{Haken197577}, and intensity fluctuations characterized by correlation functions\cite{PhysRevA.47.1431, MH10}.  

In good-cavity optical lasers, the atomic polarization (proportional to $J_\perp$) can be adiabatically eliminated and the relaxation oscillations are associated with the flow of energy back and forth between the gain inversion  (proportional to $J_z$) and the cavity field $A$, where $J_z$, $J_\perp$ are components of the collective Bloch vector $\vec{J}$ describing the atomic gain medium.  In contrast, in a bad-cavity laser, the cavity field can be adiabatically eliminated, and the oscillations are driven by the coupling of $J_\perp$ and $J_z$.  Here, we will measure and infer not only the light field $A(t)$, but also the atomic degrees of freedom $J_\perp(t)$ and $J_z(t)$ using non-demolition cavity-aided measurements\cite{CBS11, CBWtheory12} to give the complete picture of the dynamics of relaxation oscillations in a bad-cavity laser.

We will also consider the effects on the laser's amplitude stability of non-ideal repumping through multiple intermediate states. Intermediate repumping states were not included in previous simple theoretical models\cite{MYC09}, but are present in most actual realizations. In addition, we demonstrate that the cavity frequency tuning in response to the distribution of atomic population among various ground states can be used to suppress or enhance relaxation oscillations in the Raman transition configuration or other configurations with atomic transitions near-detuned from the lasing mode. As evidence, we show stabilization of $J_z$, $J_\perp$, and $A$ similar to observations of the suppression of relaxation oscillations in good-cavity lasers\cite{SDW65}. The cavity frequency tuning mechanism is related to other applications of cavity feedback including the creation of nonlinearities for generating spin-squeezed atomic ensembles \cite{LSV10}, cavity cooling and amplification in atomic and opto-mechanical systems \cite{KV08}, and the control of instabilities in gravitational wave detectors \cite{JBZ09}. 

\begin{figure}[t]
\includegraphics[width=3.3in]{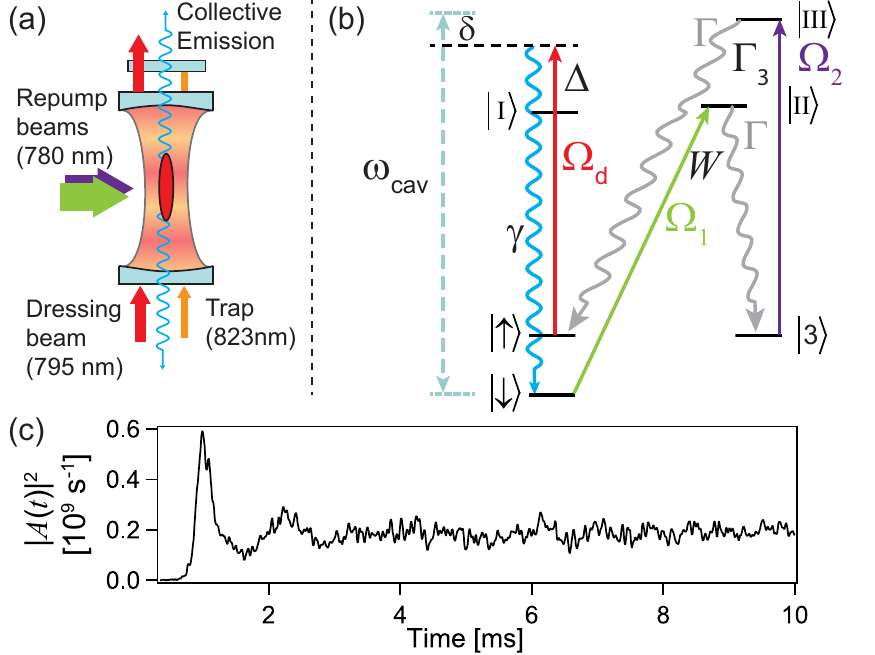}
\caption{ (a), (b) Physical setup and energy level diagram. The trapping light (orange) and Raman dressing laser (red, power $\propto \Omega^2_d$) are injected along the cavity axis. The repumping light (purple, green) is applied perpendicular to the cavity axis. The emitted optical laser light (blue) is nearly resonant with the cavity mode (dashed lines) detuned from $\omega_{\mathrm{cav}}$ by $\delta$. The repumping is accomplished through a pair of two-photon transitions through intermediate optically excited states $\ket{\mathrm{II}}$ and $\ket{\mathrm{III}}$ with incoherent decay rates $\Gamma$.  We individually control the two-photon rates $W$ and $\Gamma_3$ with the repumping laser powers $\propto \Omega_{1,2}^2$.  $\ket{3}$ represents other metastable ground states besides the laser levels.  (c) Example emitted laser photon flux $|A(t)|^2$ versus time showing spiking and relaxation oscillations at turn-on.
}
\label{ExptSetup}
\end{figure}

Our experimental system consists of a quasi-steady-state Raman laser described in Fig. \ref{ExptSetup} and in Ref. \cite{BCW12}. The laser uses $N=1\times10^6$ to $2\times10^6$ $^{87}$Rb atoms as the gain medium. The atoms are trapped and laser cooled into the Doppler-insensitive Lamb-Dicke regime (40~$\mu$K) in a 1-D optical lattice at $823$~nm formed by a standing wave in a moderate finesse $F\approx700$ optical cavity with a cavity power decay rate $\kappa/2 \pi = 11$~MHz. The single-atom cavity cooperativity parameter is $C = 8\times 10^{-3}\ll1$, and is equivalent to the Purcell factor\cite{TanjiSuzuki2011201}.

Fig. \ref{ExptSetup}b shows a simplified energy level diagram of a three level Raman laser system. The lasing transition is a spontaneous optical Raman transition with single-particle rate $\gamma$ from $\ket{\uparrow} \equiv \ket{5\,^2S_{1/2} \,F=2, m_F=0}$ to $\ket{\downarrow}\equiv\ket{5\,^2S_{1/2} \,F=1, m_F =0 }$. The decay is induced by a 795~nm dressing laser injected into the cavity non-resonantly, and detuned from the \ket{\uparrow} to $\ket{\mathrm{I}} \equiv \ket{5\,^2P_{1/2} \,F'=2}$ transition by $\Delta/2\pi = +1.1$~GHz. The atoms are incoherently repumped back to \ket{\uparrow} in two steps:  from \ket{\downarrow} to \ket{3} and then from \ket{3} to \ket{\uparrow}, at single-particle rates $W$ and $\Gto$ respectively. The third metastable ground state \ket{3} here represents the sum of all other hyperfine ground states in $^{87}$Rb. The full energy level diagram with details of the dressing and repumping lasers is provided in Ref. \cite{SOM}.  

We control $\gamma$ (typical value $60$ s$^{-1}$) using the intensity of the dressing laser.  We control the repumping rates $W$ and $\Gto$ using two 780~nm repumping lasers tuned near resonance with the $\ket{5\,^2S_{1/2} \,F=1, 2}\rightarrow \ket{5\,^2P_{3/2} \,F'=2}$ transitions. The repumping intensities are independently controlled allowing us to set the proportionality factor $r \equiv \Gto/W$. In our experiments, $W$ ranges from $10^3$ s$^{-1}$ to $10^5$ s$^{-1}$, and $r$ ranges from $0.01$ to $2$. The repumping dominates all other homogenous broadening of the $\ket{\uparrow}$ to $\ket{\downarrow}$ transition such that $\gamma_\perp \approx W/2$. The inhomogenous broadening of the transition is $\gamma_{\mathrm{in}} \approx 10^3$ s$^{-1}$. To summarize, the relevant hierarchy of rates characterizing our system is $\kappa \gg \gamma_\perp \approx W/2 \sim NC\gamma > \gamma_{\mathrm{in}} \gg \gamma$.   The rate $NC\gamma$ sets the scale for the single-particle, collectively-enhanced decay rate from \ket{\uparrow} to \ket{\downarrow}. Coupling to other transverse and longitudinal cavity modes is negligible.

The frequency of the superradiantly emitted light $\omega_{\mathrm{\gamma}}$ is set by the frequency of the dressing laser and the hyperfine splitting $\omega_{\mathrm{HF}}/2\pi=6.834$ GHz.  The detuning of light and cavity resonance frequency is $\delta = 2(\omega_{\mathrm{cav}}-\omega_{\mathrm{\gamma}})/\kappa$, normalized to the cavity half linewidth. The single particle scattering rate from the dressing laser into the cavity mode is $\Gamma_c(\delta) = C\gamma/\left(1+\delta^2 \right)$.

The cavity frequency is dispersively tuned by the atomic ensemble $\omega_{\mathrm{\mathrm{cav}}} = \omega_{\mathrm{bcav}}+ \sum_k  \alpha_k N_k$, where $\omega_{\mathrm{bcav}}$ is the bare cavity frequency and $\alpha_k$ is the cavity frequency shift for a single atom in the $k$th ground Zeeman state resulting from dispersive phase shifts of the intracavity light field.

Since the cavity frequency shift from atoms in the $F=1,2$ hyperfine states are not equal, the cavity frequency can provide a measurement of the atomic populations.  We can suddenly switch off the repumping and dressing lasers to effectively freeze the atomic populations\cite{BCW12Hybrid}. We then combine repeated non-demolition cavity frequency measurements\cite{TGJ08, SLV10, CBS11, CBWtheory12} and NMR-like rotations\cite{CBW12} to determine $J_z(t) =  \langle \frac{1}{2}\sum_{i=1}^N \left(\ket{\uparrow_i}\bra{\uparrow_i} - \ket{\downarrow_i}\bra{\downarrow_i}\right) \rangle$ and $\delta(t)$ \cite{SOM}.  We measure the amplitude of the light field emitted from the cavity $A(t)$ in heterodyne just prior to freezing the system, along with the measurement of the cavity frequency detuning $\delta$ provides an inferred value of $J_\perp = \left| \left < \hat{J}_- \right> \right|$ using the relation $A(t) = \sqrt{\Gamma_c(\delta(t))} J_\perp(t) $,  where $\hat{J}_- = \sum_{i=1}^N \ket{\downarrow_i}\bra{\uparrow_i}$. 

\begin{figure}
\includegraphics[width=3.3in]{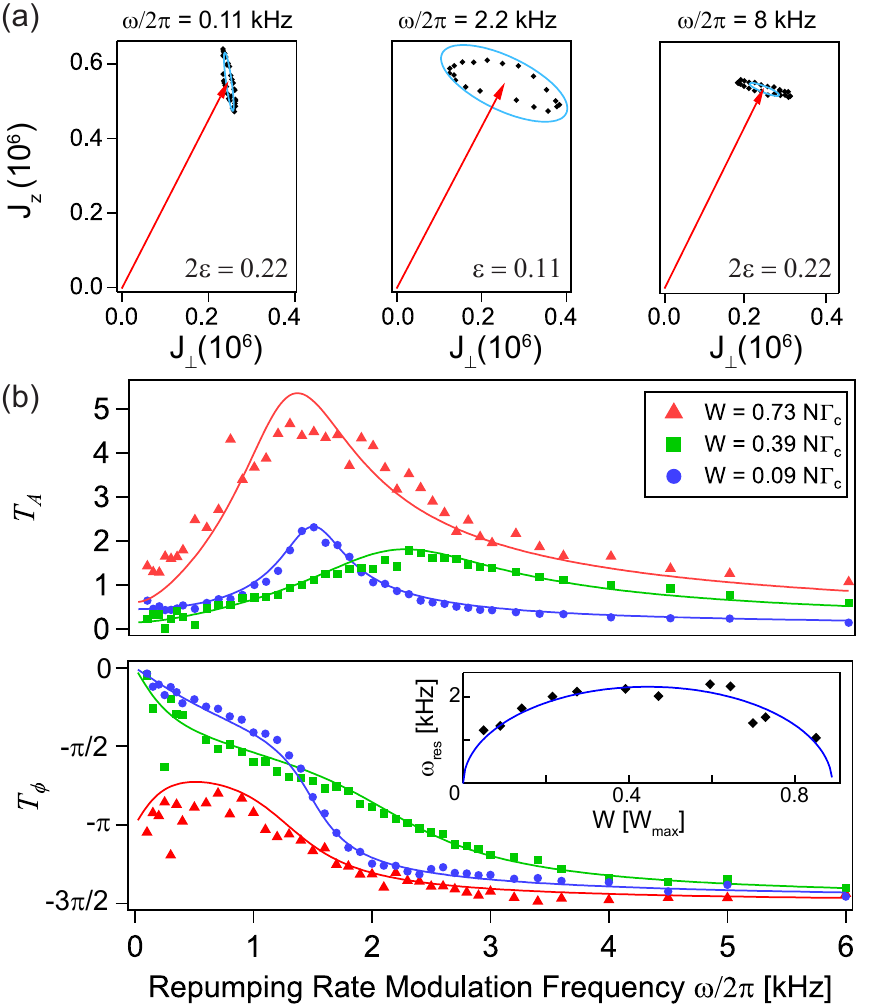}
\caption{ (a) Parametric plots of the Bloch vector components $J_z(t)$ and $J_\perp(t)$ over a single cycle of modulation of the repumping rate $W$ for modulation frequencies below, near, and above resonance or $\omega/2\pi = 0.11, 2.2, 8$ kHz, from left to right. The black points are the measured small-signal deviations about the measured steady-state Bloch vector (red arrow).  The blue curve is the predicted deviation from steady-state given the experimental parameters $N = 1.3\times 10^6$, $r= 0.71$, $\bar{\delta} = 1$, $\bar{W} = 0.35\, NC\gamma$, and $NC\gamma = 125 \times 10^3 s^{-1}$.   The modulation depth $\epsilon$ for data below and above resonance was doubled to make the response more visible.  (b) The amplitude (upper) and phase (lower) response transfer functions $T_{A}(\omega), T_{\phi}(\omega)$ of the light field for three values of the repumping rate $W$. The points are measured data, and the lines are zero free-parameter predictions of the response.  (Inset) $\omega_{\mathrm{res}}$ versus $W$ (points) and a fit to $\omega_0$ (line) showing the expected frequency dependence of the relaxation oscillations on repumping rate.
}
\label{AtomicResponse}
\end{figure}

We observe characteristic laser spiking and relaxation oscillation behavior in $|A(t)|^2$ as the laser turns on and settles to steady state (Fig. \ref{ExptSetup}c).   To  systematically study small amplitude deviations about the steady-state values $\bar{J}_{z}$, $\bar{A}$, and $\bar{J}_{\perp}$, we apply a swept sine technique, similar to Ref. \cite{Wahlstrand07}. We apply a simultaneous small amplitude modulation of the repumping rates as $W(t) = \bar{W}(1+ \epsilon~\mathrm{Re}[e^{i \omega t}]))$ and $\Gto(t) = r W(t)$. The modulation frequency $\omega$ is scanned over frequencies of order $\bar{W}$, such that $\gamma < \omega \ll \kappa$.  We then measure and infer the quantities $A(t)$, $J_\perp(t)$, and $J_z(t)$ as described earlier. 

To calculate the transfer function of the applied modulation, the measured light field amplitude $A(t)$ exiting the cavity as a function of time is fit to $A(t)= \bar{A}(1+a(\omega) \cos(\omega t + \phi_a(\omega)))$.  The normalized fractional amplitude response transfer function is $T_A(\omega)\equiv  a(\omega)/\epsilon$ and the phase response transfer function is $T_\phi \equiv \phi_a(\omega)$. We also define the modulation frequency that maximizes $T_A(\omega)$ as the resonance frequency $\omega_{\mathrm{res}}$.  

We present the measured transfer functions and atomic responses in Figs. 2, 3, and 4, with theoretical predictions from a full model for $^{87}$Rb for quantitative comparison.  To guide the interpretation of the measurements, we present an analogous 3-level model for the system shown in Fig. 1b that captures qualitative features of the full model\cite{SOM}.  The 3-level model uses semi-classical optical Bloch equations to describe the lasing transition and the repumping process.  Since $\kappa\gg W, \gamma$, the cavity field can be adiabatically eliminated from the system of equations. Additionally, we have adiabatically eliminated the populations in the optically excited states $\ket{\mathrm{I}}, \ket{\mathrm{II}}, \ket{\mathrm{III}}$, arriving at the steady state solutions for the inversion $\bar{J}_{z}$ and collective atomic coherence $\bar{J}_{\perp}$\cite{MYC09, MEH10}. The steady state amplitude $\bar{A}$ is maximized at $W = \wpk = \frac{1}{2}N\Gamma_c(\bar{\delta})$ where $\bar{\delta}$ is the steady state cavity detuning. 

To predict relaxation oscillations and damping, we do a straightforward expansion about the steady state values $\bar{J}_{z}$, $\bar{J}_{\perp}$, and $\bar{N}_3$ as $J_z(t) \approx \bar{J}_{z}(1+\mathrm{Re}[\jmath_z(t)])$, $J_\perp(t) \approx \bar{J}_{\perp}(1+\mathrm{Re}[\jmath_\perp(t)])$, and $N_3 \approx \bar{N}_3(1+\mathrm{Re}[n_3(t)])$ and ignore terms that are second order in small complex quantities $\jmath_\perp, \jmath_z,n_3$ and repumping modulation amplitude $\epsilon$ \cite{PhysRevA.47.1431,siegman86,SOM}.   The coupled quadratures  $\jmath_z$ and $\jmath_\perp$ respond like the two coupled quadratures of a harmonic oscillator, slightly modified by the presence of the intermediate repumping state \ket{3}.  In the limit of ideal repumping ($r \rightarrow \infty$) as is considered in Ref. \cite{MYC09}, we can recast the equations as two uncoupled, second order differential equations

\begin{equation}
\ddot{\jmath}_{z, \perp}+2 \gamma_0 \dot{\jmath}_{z, \perp}+ \omega_0^2 \jmath_{z, \perp} = D_{z, \perp}(\omega)\epsilon e^{i \omega t}.
\label{response}
\end{equation}

\noindent When $\bar{\delta} = 0$, the damping rate $\gamma_0 = \bar{W}/2$ is set by the damping of the transverse component $\jmath_\perp$ caused by single-particle wave function collapse associated with the repumping.  The natural frequency $\omega_0 = \sqrt{\bar{W}(NC\gamma-\bar{W})}= \sqrt{2} \bar{J}_\perp C\gamma$ is set by the steady-state rate of converting collective transverse coherence into inversion $ \bar{J}_\perp^2 C\gamma$, normalized by the total steady state coherence $\bar{J}_\perp$. 

The responses of the two quadratures to the modulation are different because the effective drives are different with $D_\perp(\omega) = \frac{\bar{W}}{2}(NC\gamma - 2\bar{W} - i\omega)$ and  $D_z(\omega)  =(NC\gamma-\bar{W}) (\bar{W}+i \omega) $. Note that the magnitude and phase of the drives change with the modulation frequency and repumping rate, even as the modulation depth $\epsilon$ remains constant.

We show this driven oscillator response in Fig. \ref{AtomicResponse}a with the measured and predicted parametric plot of $J_z$ and $J_\perp$ at three different applied modulation frequencies, with repumping near $\bar{W} =\wpk$.   Although the characteristic frequencies and rates of the atomic oscillator do not change, the differing drives lead to a change in the phase relationship between the response of the two quadratures. We believe the discrepancy with theory in the center panel of Fig.~\ref{AtomicResponse}a is the beginning of nonlinearity in the system as it responds beyond the small perturbation regime near resonance.

In Fig. \ref{AtomicResponse}b, we focus on the light field's transfer functions \TA~and \Tph.  Data for three different average repumping rates $\bar{W}$ are shown.  The data displays the features of the simple 3-level model, namely increased damping with $\bar{W}$, $\omega_0$ scaling with $\bar{W}$ (inset),  the $270^\circ$ phase shift of $T_\phi$ at high modulation frequencies, the small response near $\omega = 0$ and $\bar{W} = W_{pk}$ caused by the cancellation in the drive term $D_\perp(\omega)$, and finally the phase reversal of the response near $\omega=0$ going from below to above $W_{pk}$. The data also quantitatively agrees with the displayed theory calculated for the full model including all $^{87}$Rb levels. We suspect the deviation for $\bar{W} = 0.73 N\Gamma_c$ is a result of a systematic error in measuring the total atom number.

\begin{figure}[t]
\includegraphics[width=3.3in]{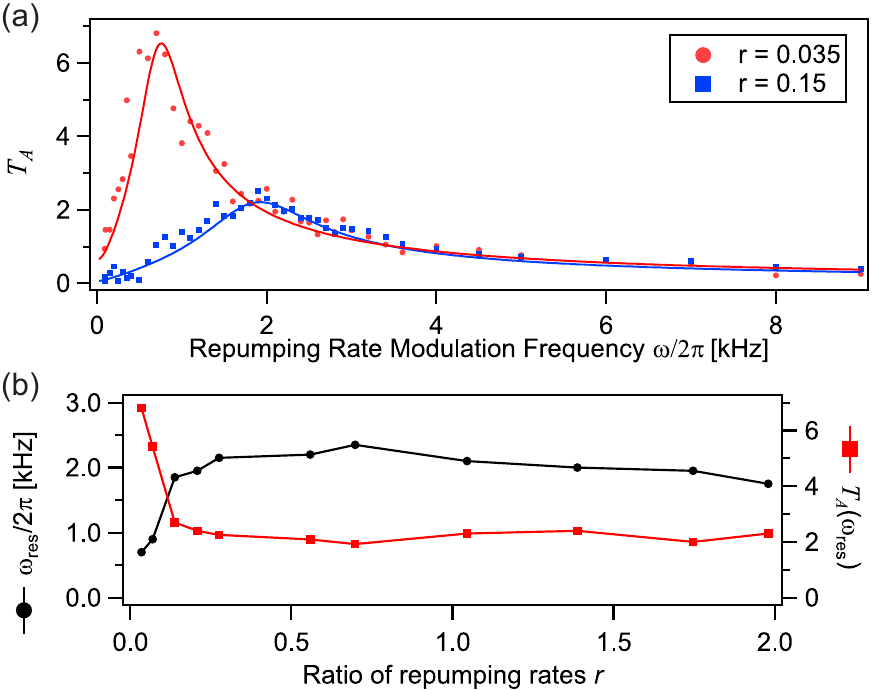}
\caption{ Effects of finite ratio of repumping rate $r$ (a) Comparison at two values of $r$ of \TA versus modulation frequency. The points are measured data in good agreement with the zero-free parameter fit (lines). (b) Plot of the resonance frequency $\omega_{\mathrm{res}}$ (black) and the peak value $T_A(\omega_{\mathrm{res}})$ (red) versus $r$.
}
\label{RSat}
\end{figure}

In this work, the repumping ratio $r \not= \infty$, and the theory must be extended to quantitatively describe the data. The physical effect of finite $r$ is that population builds up in \ket{3}. The ratio of steady state populations is simply $\bar{N}_3/\bar{N}_\downarrow = 1/r$.  As a result of a non-negligible $\bar{N}_3$, the natural frequency is slightly modified as $\omega_0 = \sqrt{\frac{r}{1+r}\bar{W}(NC\gamma-\bar{W})}$. The effective damping in the presence of a harmonic drive at frequency $\omega$ is $\gamma_0 = \frac{\bar{W}}{2}\frac{r^2}{(1+r)(1/2+r)}+\frac{r(NC\gamma - \bar{W})}{2(1+r)(1/2+r)} - \frac{\omega^2}{\bar{W}(1+r)}$.  The frequency dependent term results from the additional phase shift introduced into the oscillating system as a result of time spent in \ket{3}. Despite the frequency-dependent reduction of the damping, as long as $\gamma_0 > 0$ near $\omega=\omega_0$, the system will remain stable.  We can experimentally observe a reduction in damping as $r \rightarrow 0$, shown in Fig. \ref{RSat}.  From the form of  $\omega_0$ and $\gamma_0$, we expect to see the resonance frequency sharply decrease, and an increase in the peak relaxation oscillation amplitude, as $r \rightarrow 0$, which we observe in Fig. \ref{RSat}b. 

\begin{figure}[t]
\includegraphics[width=3.3in]{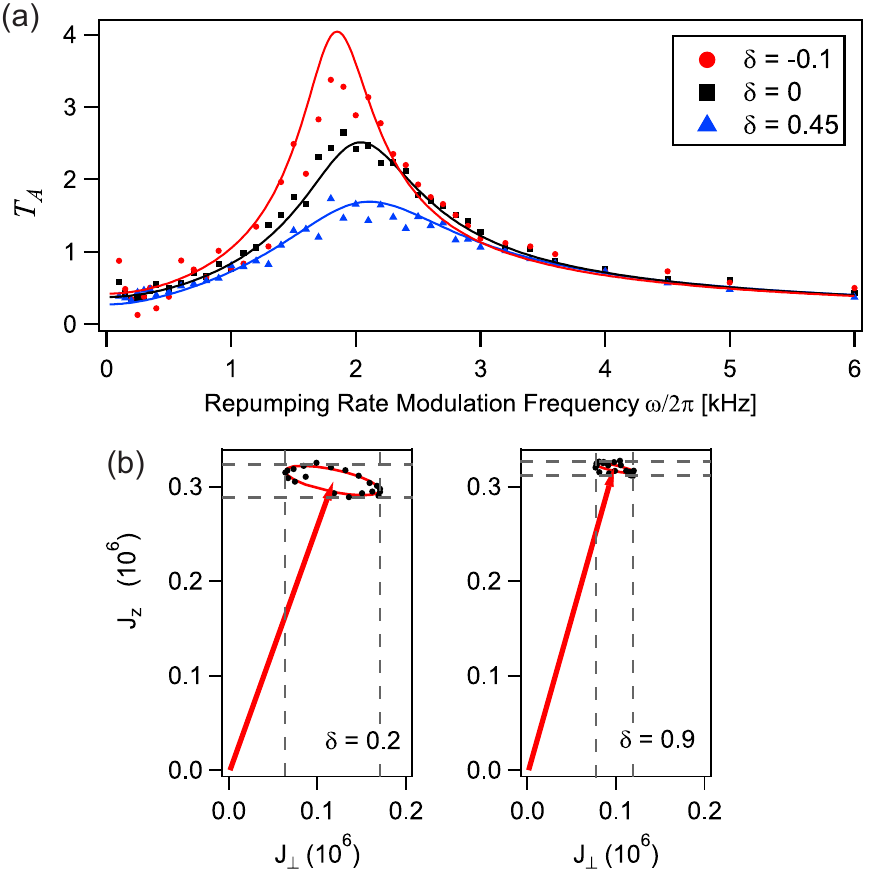}
\caption{ Evidence of negative and positive cavity feedback. (a) Amplitude transfer functions of the emitted electric field for three detunings from cavity resonance. The data points are the average of 4 experimental trials. The lines are fitted transfer functions with N as a free parameter. (b) Cavity damping of collective atomic degrees of freedom. The response going from $\bar{\delta} = 0.2$ (left) where the $\gamma_0$ is small to $\bar{\delta} = 0.9$ (right) where the system is expected to be critically damped.  The red lines are sinusoidal fits to the data (circles).  The dashed lines highlight the damping in both $J_\perp$ and $J_z$.
}
\label{CavityDamp}
\end{figure}

To understand the dynamic tuning of the cavity resonance frequency  $\omega_{\mathrm{cav}}$ in response to changes in the atomic populations, we consider the case $r\rightarrow \infty$ and $\bar{\delta} \not=0$. We also assume the cavity frequency is tuned by the atomic inversion $J_z$ as $\alpha = \alpha_\downarrow =-\alpha_\uparrow>0$.  The dynamic cavity tuning then modifies the damping rate as $\gamma_0 = \frac{\bar{W}}{2}\left(1 +2 \alpha \bar{\delta} \left(\frac{N}{1+\bar{\delta}^2}-\frac{\bar{W}}{C\gamma}\right)\right)$. The dispersive tuning of the cavity frequency can act as either positive or negative feedback on the oscillations of $\jmath_{\perp,z}$ for $\bar{\delta} < 0$ and $>0$ respectively.   As an example, in the case of negative feedback, if the inversion $J_z$ decreases, the cavity tunes away from resonance with the Raman transition, reducing the superradiant emission from \ket{\uparrow} to \ket{\downarrow}, and allowing the repumping to restore the inversion more quickly. We observe both positive and negative feedback in the measured transfer function $T_A(\omega)$ and the atomic responses $J_z(t)$ and $J_\perp(t)$ as shown in Fig. \ref{CavityDamp}.

We have studied the dynamics of the polarization, inversion, and field of an optical laser operating deep in the bad-cavity regime. We have shown that dispersive cavity frequency tuning can suppress or enhance relaxation oscillations. Having experimentally validated our model for optical lasers in the extreme bad-cavity regime, future work can now extend the formalism to realistic models of proposed ultrastable lasers using ultranarrow atomic transitions in atoms such as Sr and Yb\cite{MYC09}.  In the future, it should be possible to directly monitor $J_\perp$ using techniques similar to those presented here to monitor $J_z$. Further studies of the nonlinear dynamics of the extreme bad-cavity laser system will include investigations of chaos\cite{Haken197577} and squeezed light generation\cite{LCRG95}.

The authors acknowledge helpful discussions with D. Meiser, M. J. Holland, J. Ye, D. Z. Anderson, and S. T. Cundiff. This work was funded by NSF PFC, NIST, ARO, and DARPA QuASAR. J.G.B. acknowledges support from NSF GRF, and Z.C. acknowledges support from A*STAR Singapore.

\end{bibunit}

\begin{bibunit}
\title{Relaxation oscillations, stability, and cavity feedback in a superradiant Raman laser: supporting online material}
\author{Justin G. Bohnet}
\author{Zilong Chen}
\author{Joshua M. Weiner}
\author{Kevin C. Cox}
\author{James K. Thompson}
\affiliation{JILA, NIST and Department of Physics, University of Colorado, Boulder, Colorado 80309-0440, USA }


\maketitle
\subsection{Cavity details, full energy level diagram, dressing, and repumping laser scheme.}
The optical cavity has a mode waist of $71~ {\rm\mu m}$, mirror separation of $1.9~{\rm cm}$, and a finesse of $F= 700$. The atoms are trapped by a 1-D intracavity optical lattice at 823 nm and laser cooled to approximately 40~$\mu$K.  The sub-wavelength localization of the atoms along the cavity-axis ensures that the atoms are in the Lamb-Dicke regime along this direction.  However, the atoms are not in the Lamb-Dicke regime with respect to motion transverse to the cavity axis.

\begin{figure}
\includegraphics[width=3.5in]{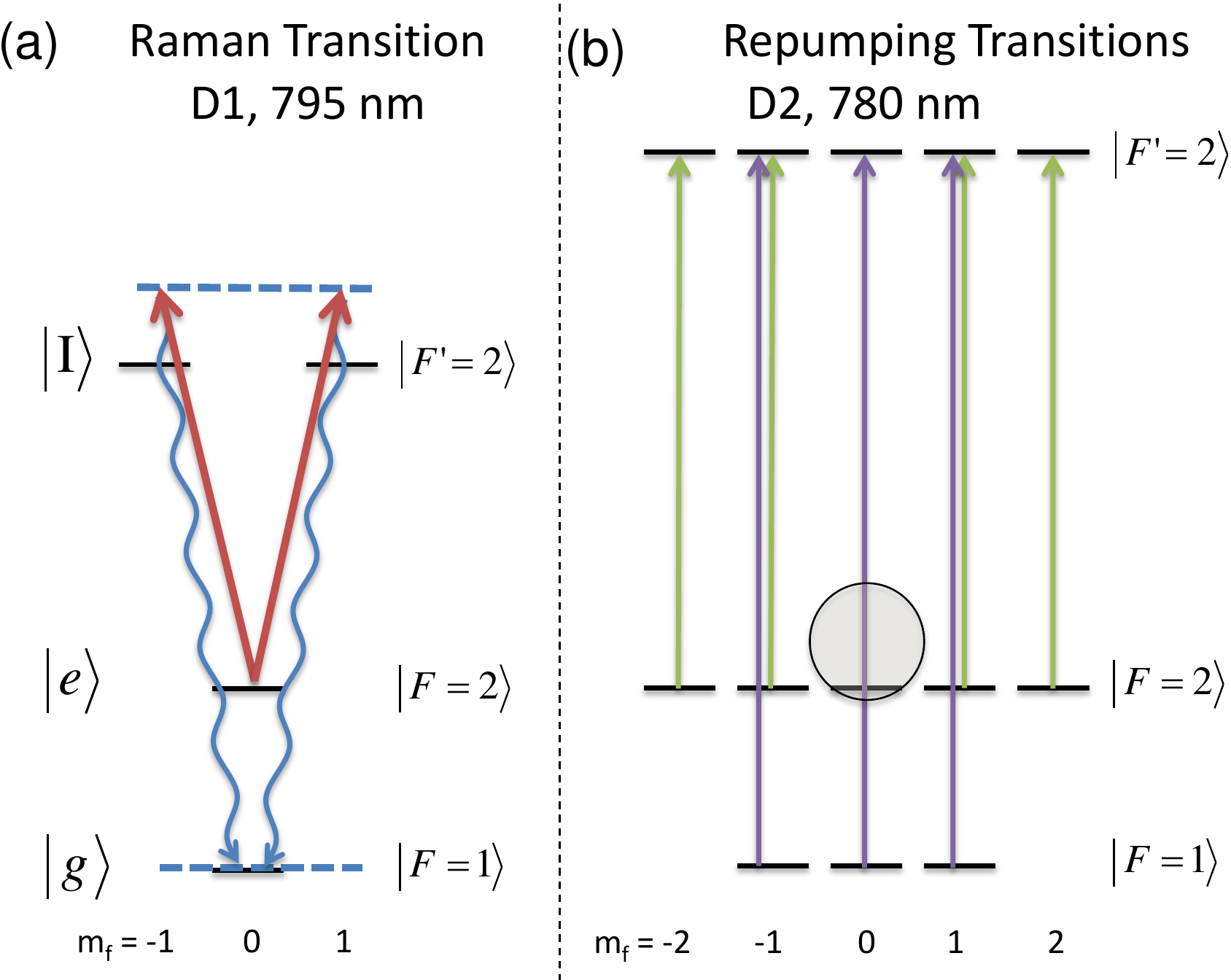}
\caption{(a) The energy level diagram for the D1 Raman transition used for lasing.  The linearly polarized Raman dressing laser is shown in red and the superradiantly emitted light in blue.  The cavity mode resonance frequency $\omega_{cav}$ is denoted with a blue dashed line.  With the quantization axis defined by a 2.7~G magnetic field oriented along the cavity axis, the linearly polarized light is a linear combination of $\sigma_+$ and $\sigma_-$ polarizations. (b) The energy level diagram for the D2 repumping beams F2 (green) and  F1 (purple). The dark state with respect to the repumping lasers is labeled with a gray circle and corresponds to \ket{\uparrow}.}
\label{EnergyLevelDiagram}
\end{figure}

We first note that the energy levels, dressing laser, repumping lasers, and relative frequency tunings are the same as the primary configuration presented in Ref. \cite{BCW12}.  We include the details here again for clarity.  The $ \ket{\downarrow} \equiv \ket{5\,^2S_{1/2} \,F=1, m_F = 0}$ and $ \ket{\uparrow}\equiv \ket{5\,^2S_{1/2} \,F=2, m_F = 0} $ hyperfine ground states of $^{87}$Rb form the basis of our Raman laser (energy level diagram shown in Fig. 5).  The \ket{\uparrow} state is dressed with an optical Raman laser at $795$ nm that induces an optical decay to \ket{\downarrow}, with the cavity tuned to be resonant or near-resonant with the emitted light.  The intensity of the dressing beam and the detuning by $\Delta/2\pi = +1.1$ GHz  from the optically excited intermediate states $\ket{\mathrm{I}_\pm}\equiv\ket{5\,^2P_{1/2} \,F'=2, m'_F = \pm1}$ set the single-atom scattering rate into all of free space $\gamma$.  The induced single-atom scattering rate into the cavity mode is $\Gamma_c = \frac{C\gamma}{1+\bar{\delta}^2}$ where $\bar{\delta}$ is the detuning of the cavity resonance frequency from the emitted light in units of the cavity half-linewidth, and $C$ is the single particle cooperativity parameter of cavity QED\cite{CBS11}, equivalent to a Purcell factor\cite{TanjiSuzuki2011201}. The branching ratio for decay to $\ket{\downarrow}$ are included in the definition of $C$.

The quantization axis is set by a $2.7~\text{G}$ magnetic field along the cavity axis. The Raman dressing laser is injected non-resonantly along the cavity axis (experimental setup shown in Fig. 6a).   For this quantization axis, the linearly polarized Raman dressing laser is an equal combination of $\sigma^+$ and $\sigma^-$ light. Constructive interference between the two decay paths from $\ket{\uparrow}$ to $\ket{\downarrow}$ through the two states $\ket{\mathrm{I}_\pm}$ leads to enhancement of light emission with linear polarization rotated $90^{\circ}$ from the polarization of the dressing laser.  Conversely,  emission of light into the cavity with the same polarization as the dressing laser is highly suppressed by destructive interference of the two decay paths.

Two repumping lasers control the rate out of \ket{\downarrow} and back into \ket{\uparrow}.   The laser powers are independently set using acoustic optic modulators (AOMs) 1 and 2 shown in Fig. 6a.  Both repumpers are $\pi$-polarized, applied perpendicular to the cavity, and separately tuned near the $\ket{5\,^2S_{1/2} \,F= 1,2 }$ to $\ket{5^2P_{3/2}, F'=2}$ transitions.  The F1 repumper moves atoms primarily from the ground $\ket{5\,^2S_{1/2}  F=1, m_F}$ states to the ground $\ket{5\,^2S_{1/2}  F=2, m_F}$ states, and sets the scattering rate \Goz out of \ket{\downarrow}.  The F2 repumper pushes population to \ket{\uparrow} as the dipole matrix element for the transition $|5^2S_{1/2}, F=2, m_f = 0\rangle \rightarrow |5^2P_{3/2}, F'=2, m_f' = 0\rangle$ is zero.  We quantify the F2 repumping rate by calculating the single particle scattering rate for an atom in $|5^2S_{1/2}, F=2, m_f = 1\rangle$ state as \Gto. We define the pump ratio $r \equiv \frac{\Gamma_3}{W}$ to quantify the degree of population buildup outside of the two level manifold $\ket{\uparrow}$ and $\ket{\downarrow}$, with the ideal repumping case being $r \rightarrow \infty$.  Though $\Gamma_{3}$ is not perfectly equivalent to $\Gamma_{3\uparrow}$ from the three level model, it allows us to define $r$ in an analogous way.  Population will still remain in the states $|5^2S_{1/2}, F=1, m_f = \pm1\rangle$ even as $r\rightarrow \infty$, unlike in the ideal three-level model. This why the output power correction factor $R$ does not asymptote to 1 as $r \rightarrow \infty$.  Also note that unlike the simple three-level model, here, $W$ includes Rayleigh scattering back to \ket{\downarrow}, contributing damping of the coherence $J_\perp$ without affecting the inversion $J_z$.

\subsection{Driven harmonic oscillator model}

Here we provide the three-level model used to arrive at the driven harmonic oscillator model in the main text, shown in Fig. 1b, and describe the extension to the theory used to model the 8 level $^{87}$Rb system. 

We start by deriving the equations of motion for the expectation values of the operators associated with the cavity mode and the reduced density matrix describing the atomic system using the master equation approach of Ref. \cite{MYC09}.  The coherent coupling of the cavity mode and the two-level subsystem is described by the standard Jaynes-Cummings Hamiltonian $\hat{H}_{JC}$\cite{MYC09}.  The atom-field coupling is described by the coupling parameter $g$, such that the vacuum-Rabi splitting is $2g$.  The evolution of the density matrix  $\hat{\rho}$ is calculated using 

\begin{equation}
\frac{d \hat{\rho}}{d\,t}= \frac{1}{\imath \hbar} [ \hat{H}_{JC}, \hat{\rho}] + \hat{\mathcal{L}}_\kappa+ \hat{\mathcal{L}}_{\downarrow 3}+ \hat{\mathcal{L}}_{3\uparrow}
\end{equation}

\noindent where Liouvillian operators have been added to describe non-Hamiltonian dynamics.  The cavity damping at power decay rate $\kappa$ is included via the term $\hat{\mathcal{L}}_\kappa$,  identical to that defined in Ref. \cite{MYC09}.  The incoherent repumping is also included similarly to Ref. \cite{MYC09}, however, the atoms in \ket{\downarrow} spontaneously transition at rate $W$ first to the intermediate state \ket{3}, then back to state \ket{\uparrow}.  The first leg of the single particle repumping is included via the operator

\begin{equation}
\hat{\mathcal{L}}_{\downarrow 3} = -\frac{W}{2}\sum^N_{i=1}(\hat{\sigma}^i_{\downarrow3}\hat{\sigma}^i_{3\downarrow}\hat{\rho} + \hat{\rho}\hat{\sigma}^i_{\downarrow3}\hat{\sigma}^i_{3\downarrow} -2\hat{\sigma}^i_{3\downarrow}\hat{\rho}\hat{\sigma}^i_{\downarrow3})
\end{equation}

\noindent where the sum is over the $N$ atoms and the single particle operators for the ith atom are defined as $\hat{\sigma}^i_{jk}\equiv \ket{j}\bra{k}$ with $j, k \in \{\downarrow, \uparrow, 3 \}$.  The single particle decay at rate $\Gamma_{3\uparrow}$ from \ket{3} to \ket{\uparrow} is described by the operator $\hat{\mathcal{L}}_{\downarrow 3}$, the same as the above with appropriate change of state labels and letting $W\rightarrow \Gamma_{3\uparrow}$.  We have neglected the small spontaneous decay rate between \ket{\uparrow} and \ket{\downarrow} due to single-particle scattering into free space, justified in the limit of our experiment for which $\gamma\ll W, N C \gamma$.

From the density matrix, we can find equations of motion for the expectation value of the collective operators defined in the main text $J_\perp$, $J_z$,  the population in \ket{3} $N_3\equiv \langle \sum_{i=1}^N \hat{\sigma}_{33}^i \rangle$, and the field annihilation operator $\hat{c}\equiv \langle{\hat{c}}\rangle$ using $\ddt\expec{\hat{\mathcal{O}}} = \mathrm{Tr}[\hat{\mathcal{O}}\frac{d\hat{\rho}}{d\mathrm{t}}]$. These form a closed set of first order non-linear differential equations, if one assumes no entanglement between the atomic degrees of freedom and the cavity mode such that expectation values of products can be factorized into products of expectation values.

The cavity field can be adiabatically eliminated from the set of coupled equations, assuming that the laser is operating deep in the bad cavity limit, where the decay rate $\gamma_\perp$ of the atomic coherence $J_\perp$ is much less than the cavity power decay rate $\kappa$ \cite{MYC09}.  Laser cooling and trapping the atoms minimizes single particle decoherence, so the atomic coherence decay $\gamma_\perp \approx W/2$ as it is  dominated by the decoherence from the repuming in $\hat{\mathcal{L}}_{\downarrow 3}$ introduced above.  After elimination of the cavity field we find

\begin{equation}
\ddt J_z =  (\Gamma_{3\uparrow}-W/2) \frac{N_3}{2} + \frac{W}{2}(N/2 - J_z) - \frac{C \gamma}{1+\bar{\delta}^2} \Jperp
\label{eqn:Jz2lvl}
\end{equation}

\begin{equation}
\ddt \Jperp = -W \Jperp  + \frac{2 C \gamma}{1+\bar{\delta}^2} J_z \Jperp
\label{eqn:Jperp2lvl}
\end{equation}

\begin{equation}
\ddt N_3 = - \left(\Gamma_{3\uparrow} + W/2 \right) N_3 + W \left(N/2 - J_z\right).
\label{eqn:N32lvl}
\end{equation}

Setting the derivatives to zero, we solve for the steady state values $\bar{N}_3$, $\bar{J}_z$, and $\bar{J}_\perp$.  We then do a linearization of the equations about the steady state solutions by substituting in these equations small time dependent oscillations about the steady state values, $J_z(t) = \bar{J}_{z}(1+\mathrm{Re}[j_z(t)])$, $J^2_\perp(t) = \bar{J}^2_{\perp}(1+\mathrm{Re}[j_\perp^2(t)])$, and $N_3 = \bar{N}_{3}(1+\mathrm{Re}[n_3(t)])$. We include a drive on the system through a modulation of the repumping rate $W(t) = \bar{W}(1+\epsilon \mathrm{Re}[e^{i \omega t}])$ and $\Gamma_{3\uparrow}(t) = r W(t)$. We also allow for dynamic tuning of the cavity mode, modeling an off-resonant dispersive cavity shift, by letting $\bar{\delta} \rightarrow \bar{\delta} + \alpha \bar{J}_z\mathrm{Re}[j_z(t)]$.  We ignore terms beyond first order in small quantities $j_z(t), j_\perp(t), n_3(t)$, and $\epsilon$.  After simplification, we find the set of three first order coupled equations

%
%
\begin{widetext}
\begin{equation}
\ddt j_z = \left(\frac{N\Gamma_c - \bar{W}}{2r+1}\right) \left( \frac{n_3}{4}(2r-1) - 2r  j_\perp - \epsilon r \right) - \epsilon \bar{W}/4 - j_z \left( \frac{\bar{W}}{2} - \frac{r\bar{W}}{\Gamma_c(1+2r)}(N\Gamma_c - \bar{W}) \frac{2 \alpha \bar{\delta}}{1+\bar{\delta}^2} \right)
\end{equation}

\begin{equation}
\ddt j_\perp = - \epsilon \frac{\bar{W}}{2}  +  j_\perp  \bar{W}+ j_z \bar{W} \left(1 - \frac{\bar{W}}{C\gamma} 2 \alpha \bar{\delta} \right)
\end{equation}

\begin{equation}
\ddt n_3 = \frac{(1+2r)}{4}\frac{\bar{W}}{\bar{W}-N\Gamma_c}\left( n_3 (N\Gamma_c-\bar{W}) +\epsilon \bar{W} + j_z 2 \bar{W} \right) 
\end{equation}
\end{widetext}

These equations can be reduced to a single third order differential equation for $j_z$ or $j_\perp$.  Terms with a coefficient of $\epsilon$ result from the modulation of the repumping rates, so we define the sum of the epsilon terms as the drive $D_{z,\perp}$.  Then we write the equations in a form suggestive of a damped harmonic oscillator as

\begin{equation}
\frac{1}{(1+r) \bar{W}} \dddot{j}_{z, \perp} + \ddot{j}_{z, \perp}+2 \gamma_0 \dot{j}_{z, \perp}+ \omega_0^2 j_{z, \perp} = D_{z, \perp}(\omega)\epsilon e^{i \omega t}.
\label{responsefull}
\end{equation}

\noindent In the limit $r\rightarrow \infty$, we have exactly the equations for a damped harmonic oscillator. At finite $r$, the third order derivative adds an additional roll off in the response at frequencies greater than $(1+r)\bar{W}$, so that the third state's effect on the frequency response is analogous to a low pass filter.  Including additional intermediate states increases the order of the effective low pass filter.  However, if all of the intermediate states also have their optical pumping modulated, as is done in our model and experiment,  then the effective drive terms $D_{z, \perp}$ also acquire higher order terms in frequency. In this case, the additional roll off of the response with frequency is balanced by increases in the drive amplitude with frequency.

To extend this three level model to quantitatively describe the $^{87}$Rb Raman system used for the experiment, we consider both the additional states involved in repumping between \ket{\uparrow} and \ket{\downarrow} and the two photon Raman lasing transition.

To reduce the Raman transition to an effective two state model, the intermediate state in the two photon Raman transition from $\ket{\uparrow}$ to $\ket{\downarrow}$ is adiabatically eliminated\cite{BPM07} to arrive at a 2-photon coupling constant $g_2$  which can be written in terms of the cooperatively parameter, the cavity decay rate and the  total Raman scattering rate $\gamma$ as  $g_2 = \sqrt{C \kappa \gamma}/2$.  This two-photon coupling then enters as the new coupling constant in the two-level Jaynes-Coupling Hamiltonian $\hat{H}_{JC}$.  Additionally, the adiabatic elimination of the intermediate state produces the dispersive tuning of the cavity mode as well as differential AC stark shifts which are absorbed into the definition of the Raman transition frequency.

The optical pumping from $\ket{\downarrow}$ to $\ket{\uparrow}$ through the many intermediate hyperfine states in $^{87}$Rb is included via equivalent Liouvillian terms as was done above.  Since only two lasers are used, it is sufficient to parameterize the scattering rates from these lasers for the \ket{\downarrow} and \ket{ F=2, m_F=1} ground states which we label $W$ and $\Gamma _{3}$, with close though imperfect analogy to the three-level $W$ and $\Gamma_{3\uparrow}$ scattering rates.  All other scattering rates are then scaled to these rates by the appropriate ratios of atomic matrix elements and optically excited state branching ratios.  The analogy of these rates with those of the simple 3-level model is inexact because the original three-level model did not allow for Rayleigh scattering back to \ket{\uparrow}, and for the simple fact that the repumping process was one-way, where as in a real system, it is possible to scatter back and forth between multiple ground states.  

Together with the effective two-level Janyes-Cummings Hamiltonian above, we then find a set of coupled first order differential equations for $J_z$, $J_\perp$, the populations of the other intermediate states, and the cavity field.  These equations are reduced via adiabatic elimination of the cavity field, and then linearized about their steady state solution.  The resulting 8 coupled equations (equal in number to the number of ground hyperfine states) are reduced to an 5th order differential equation, due to the symmetry of the Zeeman states and the repumping. As described above, the effective drives $D_{z,\perp}$ also include higher order time derivatives of the applied modulation, such that the net predicted transfer functions remain qualitatively similar to those of the original three-level model.  Predictions of this full model are shown in the figures to quantitatively compare with the data, while the qualitative features of the simple three-level model are discussed in the text for their tractability.  Additional details for both models can be found in Ref. \cite{modelpaper}.

\subsection{Response function measurement sequence}

\begin{figure*}
\includegraphics[width=7in]{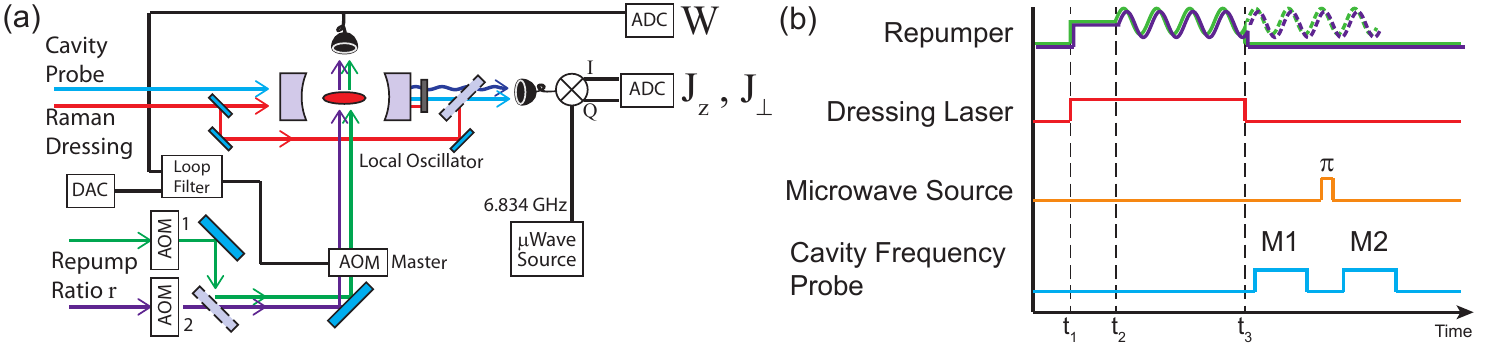}
\caption{Measurement setup and sequence timing diagram. (a) The physical setup for measuring the light field amplitude response $A(t)$ and the atomic responses $J_z(t)$, $J_\perp(t)$ to a small modulation of the repumping rate $W(t)$.  Optical beams are colored with arrows; RF and microwave signals are in black. (b)  Timing diagram for measurements. Superradiance is started at $t_1$, and the repumper $W(t)$ is modulated (with $r$ constant) starting at time $t_2$.  At $t_3$, the repumper and dressing lasers are shut off, freezing the atomic populations.  The microwave $\pi$-pulse corresponds to a $12$ $\mu$s pulse of 6.834~GHz microwaves resonant only with the ground hyperfine state transition \ket{\uparrow} to \ket{\downarrow}.  The pulse completely swaps the populations \Ne~and \Ng~of states \ket{\uparrow} and \ket{\downarrow}.  The cavity frequency probe windows M1 and M2 used to determine the dressed cavity frequency $\omega_{cav1}$ and $\omega_{cav2}$, and from which we can determine $J_z (t_3)$.}
\label{Sequence}
\end{figure*}

In the main text, we measure three quantities to characterize the response of the superradiant laser to small perturbations:  the light field amplitude $A(t)$, the magnitude of the coherence $J_\perp(t)$, and the inversion $J_z(t)$.  As shown in Fig. 6a and 6b, the perturbation is created by a small fractional  modulation of the F1 and F2 repumping laser power at frequency $\omega$ such that $W(t) = \bar{W}(1+\epsilon \cos \omega t)$ while $r(t)$ remains constant.  At time $t_3$, we shut off the repumping and dressing lasers in less than 100~ns, freezing the atomic populations\cite{BCW12Hybrid}.  The amplitude of the light just before shut off $A(t_3)$ is determined from the IQ-demodulated heterodyne signal shown in Fig. 6a.  

We then use methods similar to those in \cite{CBS11} to non-destructively probe the cavity-mode to determine the cavity resonance frequency $\omega_{cav1}$ and hence detuning from the emitted light frequency $\delta(t_3)$.   We then calculate the coherence $J_\perp(t_3)$ using the relation $A(t_3)= J_\perp(t_3) \sqrt{ C \gamma /(1+\delta^2(t_3))} $.  The probe is non-destructive in that a small fraction of the atoms are lost or Raman scattered to other states during the measurement.

The inversion $J_z(t_3)$ is determined by using a microwave $\pi$-pulse to swap the populations between \ket{\uparrow} and \ket{\downarrow}, and then measuring the cavity resonance frequency a second time $\omega_{cav2}$.  Shifts in the cavity frequency due to atoms in other states are common mode to both measurements, such that the difference between the two cavity frequency measurements is only proportional to the inversion $\omega_{cav1}-\omega_{cav2} = 2 (\alpha_\uparrow -\alpha_\downarrow) J_z(t_3) $.  The frequency tuning of the cavity resonance per atom in each state $\alpha_{\uparrow, \downarrow}$ is calculated from the known cavity geometry, atomic dipole moments, and bare cavity detuning from resonance with nearby optical atomic transitions at 795~nm.  The process is repeated for different stopping times $t_3$ in order to sample the modulation period for a set of times $t_i$.   The data $J_z(t_i)$ versus $J_\perp(t_i)$ are shown in several parametric plots in the main text. 

 In the case of the field transfer functions $T_A(\omega)$ and $T_{\phi}(\omega)$, continuous time traces $A(t)$ are averaged over several trials at the same modulation frequency and phase to enhance signal to noise.  The average response is then fit to $A(t) = A_s(1+a(\omega) \cos( \omega t+\phi_A(\omega))$, with transfer functions calculated as  $T_A(\omega) = a(\omega)/\epsilon$ and $T_{\phi}(\omega) = \phi_A(\omega)$.  The measurement is then repeated for different $\omega$.

\end{bibunit}


\begin{thebibliography}{35}%
\makeatletter
\providecommand \@ifxundefined [1]{%
 \@ifx{#1\undefined}
}%
\providecommand \@ifnum [1]{%
 \ifnum #1\expandafter \@firstoftwo
 \else \expandafter \@secondoftwo
 \fi
}%
\providecommand \@ifx [1]{%
 \ifx #1\expandafter \@firstoftwo
 \else \expandafter \@secondoftwo
 \fi
}%
\providecommand \natexlab [1]{#1}%
\providecommand \enquote  [1]{``#1''}%
\providecommand \bibnamefont  [1]{#1}%
\providecommand \bibfnamefont [1]{#1}%
\providecommand \citenamefont [1]{#1}%
\providecommand \href@noop [0]{\@secondoftwo}%
\providecommand \href [0]{\begingroup \@sanitize@url \@href}%
\providecommand \@href[1]{\@@startlink{#1}\@@href}%
\providecommand \@@href[1]{\endgroup#1\@@endlink}%
\providecommand \@sanitize@url [0]{\catcode `\\12\catcode `\$12\catcode
  `\&12\catcode `\#12\catcode `\^12\catcode `\_12\catcode `\%12\relax}%
\providecommand \@@startlink[1]{}%
\providecommand \@@endlink[0]{}%
\providecommand \url  [0]{\begingroup\@sanitize@url \@url }%
\providecommand \@url [1]{\endgroup\@href {#1}{\urlprefix }}%
\providecommand \urlprefix  [0]{URL }%
\providecommand \Eprint [0]{\href }%
\providecommand \doibase [0]{http://dx.doi.org/}%
\providecommand \selectlanguage [0]{\@gobble}%
\providecommand \bibinfo  [0]{\@secondoftwo}%
\providecommand \bibfield  [0]{\@secondoftwo}%
\providecommand \translation [1]{[#1]}%
\providecommand \BibitemOpen [0]{}%
\providecommand \bibitemStop [0]{}%
\providecommand \bibitemNoStop [0]{.\EOS\space}%
\providecommand \EOS [0]{\spacefactor3000\relax}%
\providecommand \BibitemShut  [1]{\csname bibitem#1\endcsname}%
\let\auto@bib@innerbib\@empty
\bibitem [{\citenamefont {Meiser}\ \emph {et~al.}(2009)\citenamefont {Meiser},
  \citenamefont {Ye}, \citenamefont {Carlson},\ and\ \citenamefont
  {Holland}}]{MYC09}%
  \BibitemOpen
  \bibfield  {author} {\bibinfo {author} {\bibfnamefont {D.}~\bibnamefont
  {Meiser}}, \bibinfo {author} {\bibfnamefont {J.}~\bibnamefont {Ye}}, \bibinfo
  {author} {\bibfnamefont {D.~R.}\ \bibnamefont {Carlson}}, \ and\ \bibinfo
  {author} {\bibfnamefont {M.~J.}\ \bibnamefont {Holland}},\ }\href {\doibase
  10.1103/PhysRevLett.102.163601} {\bibfield  {journal} {\bibinfo  {journal}
  {Phys. Rev. Lett.}\ }\textbf {\bibinfo {volume} {102}},\ \bibinfo {pages}
  {163601} (\bibinfo {year} {2009})}\BibitemShut {NoStop}%
\bibitem [{\citenamefont {Chen}(2009)}]{CHE09}%
  \BibitemOpen
  \bibfield  {author} {\bibinfo {author} {\bibfnamefont {J.}~\bibnamefont
  {Chen}},\ }\href@noop {} {\bibfield  {journal} {\bibinfo  {journal} {Chinese
  Science Bulletin}\ }\textbf {\bibinfo {volume} {54}},\ \bibinfo {pages} {348}
  (\bibinfo {year} {2009})}\BibitemShut {NoStop}%
\bibitem [{\citenamefont {Bohnet}\ \emph {et~al.}(2012)\citenamefont {Bohnet},
  \citenamefont {Chen}, \citenamefont {Weiner}, \citenamefont {Meiser},
  \citenamefont {Holland},\ and\ \citenamefont {Thompson}}]{BCW12}%
  \BibitemOpen
  \bibfield  {author} {\bibinfo {author} {\bibfnamefont {J.~G.}\ \bibnamefont
  {Bohnet}}, \bibinfo {author} {\bibfnamefont {Z.}~\bibnamefont {Chen}},
  \bibinfo {author} {\bibfnamefont {J.~M.}\ \bibnamefont {Weiner}}, \bibinfo
  {author} {\bibfnamefont {D.}~\bibnamefont {Meiser}}, \bibinfo {author}
  {\bibfnamefont {M.~J.}\ \bibnamefont {Holland}}, \ and\ \bibinfo {author}
  {\bibfnamefont {J.~K.}\ \bibnamefont {Thompson}},\ }\href@noop {} {\bibfield
  {journal} {\bibinfo  {journal} {Nature}\ }\textbf {\bibinfo {volume} {484}},\
  \bibinfo {pages} {78} (\bibinfo {year} {2012})}\BibitemShut {NoStop}%
\bibitem [{\citenamefont {Bohnet}\ \emph {et~al.}()\citenamefont {Bohnet},
  \citenamefont {Chen}, \citenamefont {Weiner}, \citenamefont {Cox},\ and\
  \citenamefont {Thompson}}]{BCW12Hybrid}%
  \BibitemOpen
  \bibfield  {author} {\bibinfo {author} {\bibfnamefont {J.~G.}\ \bibnamefont
  {Bohnet}}, \bibinfo {author} {\bibfnamefont {Z.}~\bibnamefont {Chen}},
  \bibinfo {author} {\bibfnamefont {J.~W.}\ \bibnamefont {Weiner}}, \bibinfo
  {author} {\bibfnamefont {K.~C.}\ \bibnamefont {Cox}}, \ and\ \bibinfo
  {author} {\bibfnamefont {J.~K.}\ \bibnamefont {Thompson}},\ }\href@noop {} {\
  }\Eprint {http://arxiv.org/abs/1208.1710v2} {arXiv:1208.1710v2
  [physics.atom-ph]} \BibitemShut {NoStop}%
\bibitem [{\citenamefont {Weiner}\ \emph {et~al.}(2012)\citenamefont {Weiner},
  \citenamefont {Cox}, \citenamefont {Bohnet}, \citenamefont {Chen},\ and\
  \citenamefont {Thompson}}]{WCB12}%
  \BibitemOpen
  \bibfield  {author} {\bibinfo {author} {\bibfnamefont {J.~M.}\ \bibnamefont
  {Weiner}}, \bibinfo {author} {\bibfnamefont {K.~C.}\ \bibnamefont {Cox}},
  \bibinfo {author} {\bibfnamefont {J.~G.}\ \bibnamefont {Bohnet}}, \bibinfo
  {author} {\bibfnamefont {Z.}~\bibnamefont {Chen}}, \ and\ \bibinfo {author}
  {\bibfnamefont {J.~K.}\ \bibnamefont {Thompson}},\ }\href {\doibase
  doi:10.1063/1.4773241} {\bibfield  {journal} {\bibinfo  {journal} {Appl.
  Phys. Lett.}\ }\textbf {\bibinfo {volume} {101}},\ \bibinfo {pages} {261107}
  (\bibinfo {year} {2012})}\BibitemShut {NoStop}%
\bibitem [{\citenamefont {Jiang}\ \emph {et~al.}(2011)\citenamefont {Jiang},
  \citenamefont {Ludlow}, \citenamefont {Lemke}, \citenamefont {Fox},
  \citenamefont {Sherman}, \citenamefont {Ma},\ and\ \citenamefont
  {Oates}}]{JLL11}%
  \BibitemOpen
  \bibfield  {author} {\bibinfo {author} {\bibfnamefont {Y.~Y.}\ \bibnamefont
  {Jiang}}, \bibinfo {author} {\bibfnamefont {A.~D.}\ \bibnamefont {Ludlow}},
  \bibinfo {author} {\bibfnamefont {N.~D.}\ \bibnamefont {Lemke}}, \bibinfo
  {author} {\bibfnamefont {R.~W.}\ \bibnamefont {Fox}}, \bibinfo {author}
  {\bibfnamefont {J.~A.}\ \bibnamefont {Sherman}}, \bibinfo {author}
  {\bibfnamefont {L.-S.}\ \bibnamefont {Ma}}, \ and\ \bibinfo {author}
  {\bibfnamefont {C.~W.}\ \bibnamefont {Oates}},\ }\href {\doibase
  10.1038/nphoton.2010.313} {\bibfield  {journal} {\bibinfo  {journal} {Nature
  Photon.}\ }\textbf {\bibinfo {volume} {5}},\ \bibinfo {pages} {158} (\bibinfo
  {year} {2011})}\BibitemShut {NoStop}%
\bibitem [{\citenamefont {Ludlow}\ \emph {et~al.}(2008)\citenamefont {Ludlow},
  \citenamefont {Zelevinsky}, \citenamefont {Campbell}, \citenamefont {Blatt},
  \citenamefont {Boyd}, \citenamefont {de~Miranda}, \citenamefont {Martin},
  \citenamefont {Thomsen}, \citenamefont {Foreman}, \citenamefont {Ye},
  \citenamefont {Fortier}, \citenamefont {Stalnaker}, \citenamefont {Diddams},
  \citenamefont {Le~Coq}, \citenamefont {Barber}, \citenamefont {Poli},
  \citenamefont {Lemke}, \citenamefont {Beck},\ and\ \citenamefont
  {Oates}}]{Ludlow28032008}%
  \BibitemOpen
  \bibfield  {author} {\bibinfo {author} {\bibfnamefont {A.~D.}\ \bibnamefont
  {Ludlow}}, \bibinfo {author} {\bibfnamefont {T.}~\bibnamefont {Zelevinsky}},
  \bibinfo {author} {\bibfnamefont {G.~K.}\ \bibnamefont {Campbell}}, \bibinfo
  {author} {\bibfnamefont {S.}~\bibnamefont {Blatt}}, \bibinfo {author}
  {\bibfnamefont {M.~M.}\ \bibnamefont {Boyd}}, \bibinfo {author}
  {\bibfnamefont {M.~H.~G.}\ \bibnamefont {de~Miranda}}, \bibinfo {author}
  {\bibfnamefont {M.~J.}\ \bibnamefont {Martin}}, \bibinfo {author}
  {\bibfnamefont {J.~W.}\ \bibnamefont {Thomsen}}, \bibinfo {author}
  {\bibfnamefont {S.~M.}\ \bibnamefont {Foreman}}, \bibinfo {author}
  {\bibfnamefont {J.}~\bibnamefont {Ye}}, \bibinfo {author} {\bibfnamefont
  {T.~M.}\ \bibnamefont {Fortier}}, \bibinfo {author} {\bibfnamefont {J.~E.}\
  \bibnamefont {Stalnaker}}, \bibinfo {author} {\bibfnamefont {S.~A.}\
  \bibnamefont {Diddams}}, \bibinfo {author} {\bibfnamefont {Y.}~\bibnamefont
  {Le~Coq}}, \bibinfo {author} {\bibfnamefont {Z.~W.}\ \bibnamefont {Barber}},
  \bibinfo {author} {\bibfnamefont {N.}~\bibnamefont {Poli}}, \bibinfo {author}
  {\bibfnamefont {N.~D.}\ \bibnamefont {Lemke}}, \bibinfo {author}
  {\bibfnamefont {K.~M.}\ \bibnamefont {Beck}}, \ and\ \bibinfo {author}
  {\bibfnamefont {C.~W.}\ \bibnamefont {Oates}},\ }\href {\doibase
  10.1126/science.1153341} {\bibfield  {journal} {\bibinfo  {journal}
  {Science}\ }\textbf {\bibinfo {volume} {319}},\ \bibinfo {pages} {1805}
  (\bibinfo {year} {2008})}\BibitemShut {NoStop}%
\bibitem [{\citenamefont {Chou}\ \emph {et~al.}(2010)\citenamefont {Chou},
  \citenamefont {Hume}, \citenamefont {Rosenband},\ and\ \citenamefont
  {Wineland}}]{CHR10}%
  \BibitemOpen
  \bibfield  {author} {\bibinfo {author} {\bibfnamefont {C.~W.}\ \bibnamefont
  {Chou}}, \bibinfo {author} {\bibfnamefont {D.~B.}\ \bibnamefont {Hume}},
  \bibinfo {author} {\bibfnamefont {T.}~\bibnamefont {Rosenband}}, \ and\
  \bibinfo {author} {\bibfnamefont {D.~J.}\ \bibnamefont {Wineland}},\ }\href
  {\doibase 10.1126/science.1192720} {\bibfield  {journal} {\bibinfo  {journal}
  {Science}\ }\textbf {\bibinfo {volume} {329}},\ \bibinfo {pages} {1630}
  (\bibinfo {year} {2010})}\BibitemShut {NoStop}%
\bibitem [{\citenamefont {Fortier}\ \emph {et~al.}(2007)\citenamefont
  {Fortier}, \citenamefont {Ashby}, \citenamefont {Bergquist}, \citenamefont
  {Delaney}, \citenamefont {Diddams}, \citenamefont {Heavner}, \citenamefont
  {Hollberg}, \citenamefont {Itano}, \citenamefont {Jefferts}, \citenamefont
  {Kim}, \citenamefont {Levi}, \citenamefont {Lorini}, \citenamefont {Oskay},
  \citenamefont {Parker}, \citenamefont {Shirley},\ and\ \citenamefont
  {Stalnaker}}]{PhysRevLett.98.070801}%
  \BibitemOpen
  \bibfield  {author} {\bibinfo {author} {\bibfnamefont {T.~M.}\ \bibnamefont
  {Fortier}}, \bibinfo {author} {\bibfnamefont {N.}~\bibnamefont {Ashby}},
  \bibinfo {author} {\bibfnamefont {J.~C.}\ \bibnamefont {Bergquist}}, \bibinfo
  {author} {\bibfnamefont {M.~J.}\ \bibnamefont {Delaney}}, \bibinfo {author}
  {\bibfnamefont {S.~A.}\ \bibnamefont {Diddams}}, \bibinfo {author}
  {\bibfnamefont {T.~P.}\ \bibnamefont {Heavner}}, \bibinfo {author}
  {\bibfnamefont {L.}~\bibnamefont {Hollberg}}, \bibinfo {author}
  {\bibfnamefont {W.~M.}\ \bibnamefont {Itano}}, \bibinfo {author}
  {\bibfnamefont {S.~R.}\ \bibnamefont {Jefferts}}, \bibinfo {author}
  {\bibfnamefont {K.}~\bibnamefont {Kim}}, \bibinfo {author} {\bibfnamefont
  {F.}~\bibnamefont {Levi}}, \bibinfo {author} {\bibfnamefont {L.}~\bibnamefont
  {Lorini}}, \bibinfo {author} {\bibfnamefont {W.~H.}\ \bibnamefont {Oskay}},
  \bibinfo {author} {\bibfnamefont {T.~E.}\ \bibnamefont {Parker}}, \bibinfo
  {author} {\bibfnamefont {J.}~\bibnamefont {Shirley}}, \ and\ \bibinfo
  {author} {\bibfnamefont {J.~E.}\ \bibnamefont {Stalnaker}},\ }\href {\doibase
  10.1103/PhysRevLett.98.070801} {\bibfield  {journal} {\bibinfo  {journal}
  {Phys. Rev. Lett.}\ }\textbf {\bibinfo {volume} {98}},\ \bibinfo {pages}
  {070801} (\bibinfo {year} {2007})}\BibitemShut {NoStop}%
\bibitem [{\citenamefont {Blatt}\ \emph {et~al.}(2008)\citenamefont {Blatt},
  \citenamefont {Ludlow}, \citenamefont {Campbell}, \citenamefont {Thomsen},
  \citenamefont {Zelevinsky}, \citenamefont {Boyd}, \citenamefont {Ye},
  \citenamefont {Baillard}, \citenamefont {Fouch\'e}, \citenamefont
  {Le~Targat}, \citenamefont {Brusch}, \citenamefont {Lemonde}, \citenamefont
  {Takamoto}, \citenamefont {Hong}, \citenamefont {Katori},\ and\ \citenamefont
  {Flambaum}}]{PhysRevLett.100.140801}%
  \BibitemOpen
  \bibfield  {author} {\bibinfo {author} {\bibfnamefont {S.}~\bibnamefont
  {Blatt}}, \bibinfo {author} {\bibfnamefont {A.~D.}\ \bibnamefont {Ludlow}},
  \bibinfo {author} {\bibfnamefont {G.~K.}\ \bibnamefont {Campbell}}, \bibinfo
  {author} {\bibfnamefont {J.~W.}\ \bibnamefont {Thomsen}}, \bibinfo {author}
  {\bibfnamefont {T.}~\bibnamefont {Zelevinsky}}, \bibinfo {author}
  {\bibfnamefont {M.~M.}\ \bibnamefont {Boyd}}, \bibinfo {author}
  {\bibfnamefont {J.}~\bibnamefont {Ye}}, \bibinfo {author} {\bibfnamefont
  {X.}~\bibnamefont {Baillard}}, \bibinfo {author} {\bibfnamefont
  {M.}~\bibnamefont {Fouch\'e}}, \bibinfo {author} {\bibfnamefont
  {R.}~\bibnamefont {Le~Targat}}, \bibinfo {author} {\bibfnamefont
  {A.}~\bibnamefont {Brusch}}, \bibinfo {author} {\bibfnamefont
  {P.}~\bibnamefont {Lemonde}}, \bibinfo {author} {\bibfnamefont
  {M.}~\bibnamefont {Takamoto}}, \bibinfo {author} {\bibfnamefont {F.-L.}\
  \bibnamefont {Hong}}, \bibinfo {author} {\bibfnamefont {H.}~\bibnamefont
  {Katori}}, \ and\ \bibinfo {author} {\bibfnamefont {V.~V.}\ \bibnamefont
  {Flambaum}},\ }\href {\doibase 10.1103/PhysRevLett.100.140801} {\bibfield
  {journal} {\bibinfo  {journal} {Phys. Rev. Lett.}\ }\textbf {\bibinfo
  {volume} {100}},\ \bibinfo {pages} {140801} (\bibinfo {year}
  {2008})}\BibitemShut {NoStop}%
\bibitem [{\citenamefont {McCumber}(1966)}]{M66}%
  \BibitemOpen
  \bibfield  {author} {\bibinfo {author} {\bibfnamefont {D.~E.}\ \bibnamefont
  {McCumber}},\ }\href {\doibase 10.1103/PhysRev.141.306} {\bibfield  {journal}
  {\bibinfo  {journal} {Phys. Rev.}\ }\textbf {\bibinfo {volume} {141}},\
  \bibinfo {pages} {306} (\bibinfo {year} {1966})}\BibitemShut {NoStop}%
\bibitem [{\citenamefont {Kuppens}\ \emph {et~al.}(1994)\citenamefont
  {Kuppens}, \citenamefont {van Exter},\ and\ \citenamefont
  {Woerdman}}]{PhysRevLett.72.3815}%
  \BibitemOpen
  \bibfield  {author} {\bibinfo {author} {\bibfnamefont {S.~J.~M.}\
  \bibnamefont {Kuppens}}, \bibinfo {author} {\bibfnamefont {M.~P.}\
  \bibnamefont {van Exter}}, \ and\ \bibinfo {author} {\bibfnamefont {J.~P.}\
  \bibnamefont {Woerdman}},\ }\href {\doibase 10.1103/PhysRevLett.72.3815}
  {\bibfield  {journal} {\bibinfo  {journal} {Phys. Rev. Lett.}\ }\textbf
  {\bibinfo {volume} {72}},\ \bibinfo {pages} {3815} (\bibinfo {year}
  {1994})}\BibitemShut {NoStop}%
\bibitem [{\citenamefont {Casperson}(1978)}]{C78}%
  \BibitemOpen
  \bibfield  {author} {\bibinfo {author} {\bibfnamefont {L.}~\bibnamefont
  {Casperson}},\ }\href {\doibase 10.1109/JQE.1978.1069683} {\bibfield
  {journal} {\bibinfo  {journal} {IEEE J. of Quantum Electron.}\ }\textbf
  {\bibinfo {volume} {14}},\ \bibinfo {pages} {756 } (\bibinfo {year}
  {1978})}\BibitemShut {NoStop}%
\bibitem [{\citenamefont {van Eijkelenborg}\ \emph {et~al.}(1998)\citenamefont
  {van Eijkelenborg}, \citenamefont {van Exter},\ and\ \citenamefont
  {Woerdman}}]{EEW98}%
  \BibitemOpen
  \bibfield  {author} {\bibinfo {author} {\bibfnamefont {M.~A.}\ \bibnamefont
  {van Eijkelenborg}}, \bibinfo {author} {\bibfnamefont {M.~P.}\ \bibnamefont
  {van Exter}}, \ and\ \bibinfo {author} {\bibfnamefont {J.~P.}\ \bibnamefont
  {Woerdman}},\ }\href {\doibase 10.1103/PhysRevA.57.571} {\bibfield  {journal}
  {\bibinfo  {journal} {Phys. Rev. A}\ }\textbf {\bibinfo {volume} {57}},\
  \bibinfo {pages} {571} (\bibinfo {year} {1998})}\BibitemShut {NoStop}%
\bibitem [{\citenamefont {Harrison}\ and\ \citenamefont {Biswas}(1985)}]{HB85}%
  \BibitemOpen
  \bibfield  {author} {\bibinfo {author} {\bibfnamefont {R.~G.}\ \bibnamefont
  {Harrison}}\ and\ \bibinfo {author} {\bibfnamefont {D.~J.}\ \bibnamefont
  {Biswas}},\ }\href {\doibase 10.1103/PhysRevLett.55.63} {\bibfield  {journal}
  {\bibinfo  {journal} {Phys. Rev. Lett.}\ }\textbf {\bibinfo {volume} {55}},\
  \bibinfo {pages} {63} (\bibinfo {year} {1985})}\BibitemShut {NoStop}%
\bibitem [{\citenamefont {Weiss}\ \emph {et~al.}(1988)\citenamefont {Weiss},
  \citenamefont {Abraham},\ and\ \citenamefont {H\"ubner}}]{WAH88}%
  \BibitemOpen
  \bibfield  {author} {\bibinfo {author} {\bibfnamefont {C.~O.}\ \bibnamefont
  {Weiss}}, \bibinfo {author} {\bibfnamefont {N.~B.}\ \bibnamefont {Abraham}},
  \ and\ \bibinfo {author} {\bibfnamefont {U.}~\bibnamefont {H\"ubner}},\
  }\href {\doibase 10.1103/PhysRevLett.61.1587} {\bibfield  {journal} {\bibinfo
   {journal} {Phys. Rev. Lett.}\ }\textbf {\bibinfo {volume} {61}},\ \bibinfo
  {pages} {1587} (\bibinfo {year} {1988})}\BibitemShut {NoStop}%
\bibitem [{\citenamefont {Shirley}(1968)}]{shirley:949}%
  \BibitemOpen
  \bibfield  {author} {\bibinfo {author} {\bibfnamefont {J.~H.}\ \bibnamefont
  {Shirley}},\ }\href {\doibase 10.1119/1.1974361} {\bibfield  {journal}
  {\bibinfo  {journal} {Am. J. Phys.}\ }\textbf {\bibinfo {volume} {36}},\
  \bibinfo {pages} {949} (\bibinfo {year} {1968})}\BibitemShut {NoStop}%
\bibitem [{\citenamefont {Kolobov}\ \emph {et~al.}(1993)\citenamefont
  {Kolobov}, \citenamefont {Davidovich}, \citenamefont {Giacobino},\ and\
  \citenamefont {Fabre}}]{PhysRevA.47.1431}%
  \BibitemOpen
  \bibfield  {author} {\bibinfo {author} {\bibfnamefont {M.~I.}\ \bibnamefont
  {Kolobov}}, \bibinfo {author} {\bibfnamefont {L.}~\bibnamefont {Davidovich}},
  \bibinfo {author} {\bibfnamefont {E.}~\bibnamefont {Giacobino}}, \ and\
  \bibinfo {author} {\bibfnamefont {C.}~\bibnamefont {Fabre}},\ }\href
  {\doibase 10.1103/PhysRevA.47.1431} {\bibfield  {journal} {\bibinfo
  {journal} {Phys. Rev. A}\ }\textbf {\bibinfo {volume} {47}},\ \bibinfo
  {pages} {1431} (\bibinfo {year} {1993})}\BibitemShut {NoStop}%
\bibitem [{\citenamefont {Haken}(1975)}]{Haken197577}%
  \BibitemOpen
  \bibfield  {author} {\bibinfo {author} {\bibfnamefont {H.}~\bibnamefont
  {Haken}},\ }\href {\doibase 10.1016/0375-9601(75)90353-9} {\bibfield
  {journal} {\bibinfo  {journal} {Phys. Lett. A}\ }\textbf {\bibinfo {volume}
  {53}},\ \bibinfo {pages} {77 } (\bibinfo {year} {1975})}\BibitemShut
  {NoStop}%
\bibitem [{\citenamefont {Meiser}\ and\ \citenamefont
  {Holland}(2010{\natexlab{a}})}]{MH10}%
  \BibitemOpen
  \bibfield  {author} {\bibinfo {author} {\bibfnamefont {D.}~\bibnamefont
  {Meiser}}\ and\ \bibinfo {author} {\bibfnamefont {M.~J.}\ \bibnamefont
  {Holland}},\ }\href {\doibase 10.1103/PhysRevA.81.063827} {\bibfield
  {journal} {\bibinfo  {journal} {Phys. Rev. A}\ }\textbf {\bibinfo {volume}
  {81}},\ \bibinfo {pages} {063827} (\bibinfo {year}
  {2010}{\natexlab{a}})}\BibitemShut {NoStop}%
\bibitem [{\citenamefont {Chen}\ \emph {et~al.}(2011)\citenamefont {Chen},
  \citenamefont {Bohnet}, \citenamefont {Sankar}, \citenamefont {Dai},\ and\
  \citenamefont {Thompson}}]{CBS11}%
  \BibitemOpen
  \bibfield  {author} {\bibinfo {author} {\bibfnamefont {Z.}~\bibnamefont
  {Chen}}, \bibinfo {author} {\bibfnamefont {J.~G.}\ \bibnamefont {Bohnet}},
  \bibinfo {author} {\bibfnamefont {S.~R.}\ \bibnamefont {Sankar}}, \bibinfo
  {author} {\bibfnamefont {J.}~\bibnamefont {Dai}}, \ and\ \bibinfo {author}
  {\bibfnamefont {J.~K.}\ \bibnamefont {Thompson}},\ }\href {\doibase
  10.1103/PhysRevLett.106.133601} {\bibfield  {journal} {\bibinfo  {journal}
  {Phys. Rev. Lett.}\ }\textbf {\bibinfo {volume} {106}},\ \bibinfo {pages}
  {133601} (\bibinfo {year} {2011})}\BibitemShut {NoStop}%
\bibitem [{\citenamefont {Chen}\ \emph {et~al.}()\citenamefont {Chen},
  \citenamefont {Bohnet}, \citenamefont {Weiner}, \citenamefont {Cox},\ and\
  \citenamefont {Thompson}}]{CBWtheory12}%
  \BibitemOpen
  \bibfield  {author} {\bibinfo {author} {\bibfnamefont {Z.}~\bibnamefont
  {Chen}}, \bibinfo {author} {\bibfnamefont {J.~G.}\ \bibnamefont {Bohnet}},
  \bibinfo {author} {\bibfnamefont {J.~W.}\ \bibnamefont {Weiner}}, \bibinfo
  {author} {\bibfnamefont {K.~C.}\ \bibnamefont {Cox}}, \ and\ \bibinfo
  {author} {\bibfnamefont {J.~K.}\ \bibnamefont {Thompson}},\ }\href@noop {}
  {\enquote {\bibinfo {title} {Cavity-aided non-demolition measurements for
  atom counting and spin squeezing},}\ }\Eprint
  {http://arxiv.org/abs/1211.0723} {arXiv:1211.0723 [physics.atom-ph]}
  \BibitemShut {NoStop}%
\bibitem [{\citenamefont {Statz}\ \emph {et~al.}(1965)\citenamefont {Statz},
  \citenamefont {DeMars}, \citenamefont {Wilson},\ and\ \citenamefont
  {Tang}}]{SDW65}%
  \BibitemOpen
  \bibfield  {author} {\bibinfo {author} {\bibfnamefont {H.}~\bibnamefont
  {Statz}}, \bibinfo {author} {\bibfnamefont {G.~A.}\ \bibnamefont {DeMars}},
  \bibinfo {author} {\bibfnamefont {D.~T.}\ \bibnamefont {Wilson}}, \ and\
  \bibinfo {author} {\bibfnamefont {C.~L.}\ \bibnamefont {Tang}},\ }\href
  {\doibase doi:10.1063/1.1703078} {\bibfield  {journal} {\bibinfo  {journal}
  {J. Appl. Phys.}\ }\textbf {\bibinfo {volume} {36}},\ \bibinfo {pages} {1510}
  (\bibinfo {year} {1965})}\BibitemShut {NoStop}%
\bibitem [{\citenamefont {Leroux}\ \emph {et~al.}(2010)\citenamefont {Leroux},
  \citenamefont {Schleier-Smith},\ and\ \citenamefont
  {Vuleti\ifmmode~\acute{c}\else \'{c}\fi{}}}]{LSV10}%
  \BibitemOpen
  \bibfield  {author} {\bibinfo {author} {\bibfnamefont {I.~D.}\ \bibnamefont
  {Leroux}}, \bibinfo {author} {\bibfnamefont {M.~H.}\ \bibnamefont
  {Schleier-Smith}}, \ and\ \bibinfo {author} {\bibfnamefont {V.}~\bibnamefont
  {Vuleti\ifmmode~\acute{c}\else \'{c}\fi{}}},\ }\href {\doibase
  10.1103/PhysRevLett.104.073602} {\bibfield  {journal} {\bibinfo  {journal}
  {Phys. Rev. Lett.}\ }\textbf {\bibinfo {volume} {104}},\ \bibinfo {pages}
  {073602} (\bibinfo {year} {2010})}\BibitemShut {NoStop}%
\bibitem [{\citenamefont {Kippenberg}\ and\ \citenamefont
  {Vahala}(2008)}]{KV08}%
  \BibitemOpen
  \bibfield  {author} {\bibinfo {author} {\bibfnamefont {T.~J.}\ \bibnamefont
  {Kippenberg}}\ and\ \bibinfo {author} {\bibfnamefont {K.~J.}\ \bibnamefont
  {Vahala}},\ }\href {\doibase 10.1126/science.1156032} {\bibfield  {journal}
  {\bibinfo  {journal} {Science}\ }\textbf {\bibinfo {volume} {321}},\ \bibinfo
  {pages} {1172} (\bibinfo {year} {2008})}\BibitemShut {NoStop}%
\bibitem [{\citenamefont {Ju}\ \emph {et~al.}(2009)\citenamefont {Ju},
  \citenamefont {Blair}, \citenamefont {Zhao}, \citenamefont {Gras},
  \citenamefont {Zhang}, \citenamefont {Barriga}, \citenamefont {Miao},
  \citenamefont {Fan},\ and\ \citenamefont {Merrill}}]{JBZ09}%
  \BibitemOpen
  \bibfield  {author} {\bibinfo {author} {\bibfnamefont {L.}~\bibnamefont
  {Ju}}, \bibinfo {author} {\bibfnamefont {D.~G.}\ \bibnamefont {Blair}},
  \bibinfo {author} {\bibfnamefont {C.}~\bibnamefont {Zhao}}, \bibinfo {author}
  {\bibfnamefont {S.}~\bibnamefont {Gras}}, \bibinfo {author} {\bibfnamefont
  {Z.}~\bibnamefont {Zhang}}, \bibinfo {author} {\bibfnamefont
  {P.}~\bibnamefont {Barriga}}, \bibinfo {author} {\bibfnamefont
  {H.}~\bibnamefont {Miao}}, \bibinfo {author} {\bibfnamefont {Y.}~\bibnamefont
  {Fan}}, \ and\ \bibinfo {author} {\bibfnamefont {L.}~\bibnamefont
  {Merrill}},\ }\href@noop {} {\bibfield  {journal} {\bibinfo  {journal}
  {Classical Quantum Gravity}\ }\textbf {\bibinfo {volume} {26}},\ \bibinfo
  {pages} {015002} (\bibinfo {year} {2009})}\BibitemShut {NoStop}%
\bibitem [{\citenamefont {Tanji-Suzuki}\ \emph {et~al.}(2011)\citenamefont
  {Tanji-Suzuki}, \citenamefont {Leroux}, \citenamefont {Schleier-Smith},
  \citenamefont {Cetina}, \citenamefont {Grier}, \citenamefont {Simon},\ and\
  \citenamefont {Vuleti\ifmmode~\acute{c}\else
  \'{c}\fi{}}}]{TanjiSuzuki2011201}%
  \BibitemOpen
  \bibfield  {author} {\bibinfo {author} {\bibfnamefont {H.}~\bibnamefont
  {Tanji-Suzuki}}, \bibinfo {author} {\bibfnamefont {I.~D.}\ \bibnamefont
  {Leroux}}, \bibinfo {author} {\bibfnamefont {M.~H.}\ \bibnamefont
  {Schleier-Smith}}, \bibinfo {author} {\bibfnamefont {M.}~\bibnamefont
  {Cetina}}, \bibinfo {author} {\bibfnamefont {A.~T.}\ \bibnamefont {Grier}},
  \bibinfo {author} {\bibfnamefont {J.}~\bibnamefont {Simon}}, \ and\ \bibinfo
  {author} {\bibfnamefont {V.}~\bibnamefont {Vuleti\ifmmode~\acute{c}\else
  \'{c}\fi{}}},\ }\href {\doibase 10.1016/B978-0-12-385508-4.00004-8}
  {\bibfield  {journal} {\bibinfo  {journal} {Adv. At. Mol. Opt. Phys.}\
  }\textbf {\bibinfo {volume} {60}},\ \bibinfo {pages} {201} (\bibinfo {year}
  {2011})}\BibitemShut {NoStop}%
\bibitem [{SOM()}]{SOM}%
  \BibitemOpen
  \href@noop {} {\enquote {\bibinfo {title} {See supplemental material for
  experimental and theoretical details},}\ }\BibitemShut {NoStop}%
\bibitem [{\citenamefont {Teper}\ \emph {et~al.}(2008)\citenamefont {Teper},
  \citenamefont {Vrijsen}, \citenamefont {Lee},\ and\ \citenamefont
  {Kasevich}}]{TGJ08}%
  \BibitemOpen
  \bibfield  {author} {\bibinfo {author} {\bibfnamefont {I.}~\bibnamefont
  {Teper}}, \bibinfo {author} {\bibfnamefont {G.}~\bibnamefont {Vrijsen}},
  \bibinfo {author} {\bibfnamefont {J.}~\bibnamefont {Lee}}, \ and\ \bibinfo
  {author} {\bibfnamefont {M.~A.}\ \bibnamefont {Kasevich}},\ }\href {\doibase
  10.1103/PhysRevA.78.051803} {\bibfield  {journal} {\bibinfo  {journal} {Phys.
  Rev. A}\ }\textbf {\bibinfo {volume} {78}},\ \bibinfo {pages} {051803}
  (\bibinfo {year} {2008})}\BibitemShut {NoStop}%
\bibitem [{\citenamefont {Schleier-Smith}\ \emph {et~al.}(2010)\citenamefont
  {Schleier-Smith}, \citenamefont {Leroux},\ and\ \citenamefont
  {Vuleti\ifmmode~\acute{c}\else \'{c}\fi{}}}]{SLV10}%
  \BibitemOpen
  \bibfield  {author} {\bibinfo {author} {\bibfnamefont {M.~H.}\ \bibnamefont
  {Schleier-Smith}}, \bibinfo {author} {\bibfnamefont {I.~D.}\ \bibnamefont
  {Leroux}}, \ and\ \bibinfo {author} {\bibfnamefont {V.}~\bibnamefont
  {Vuleti\ifmmode~\acute{c}\else \'{c}\fi{}}},\ }\href {\doibase
  10.1103/PhysRevLett.104.073604} {\bibfield  {journal} {\bibinfo  {journal}
  {Phys. Rev. Lett.}\ }\textbf {\bibinfo {volume} {104}},\ \bibinfo {pages}
  {073604} (\bibinfo {year} {2010})}\BibitemShut {NoStop}%
\bibitem [{\citenamefont {Chen}\ \emph {et~al.}(2012)\citenamefont {Chen},
  \citenamefont {Bohnet}, \citenamefont {Weiner},\ and\ \citenamefont
  {Thompson}}]{CBW12}%
  \BibitemOpen
  \bibfield  {author} {\bibinfo {author} {\bibfnamefont {Z.}~\bibnamefont
  {Chen}}, \bibinfo {author} {\bibfnamefont {J.~G.}\ \bibnamefont {Bohnet}},
  \bibinfo {author} {\bibfnamefont {J.~M.}\ \bibnamefont {Weiner}}, \ and\
  \bibinfo {author} {\bibfnamefont {J.~K.}\ \bibnamefont {Thompson}},\ }\href
  {\doibase doi:10.1063/1.3700247} {\bibfield  {journal} {\bibinfo  {journal}
  {Rev. Sci. Instrum.}\ }\textbf {\bibinfo {volume} {83}},\ \bibinfo {pages}
  {044701} (\bibinfo {year} {2012})}\BibitemShut {NoStop}%
\bibitem [{\citenamefont {Wahlstrand}\ \emph {et~al.}(2007)\citenamefont
  {Wahlstrand}, \citenamefont {Willits}, \citenamefont {Schibli}, \citenamefont
  {Menyuk},\ and\ \citenamefont {Cundiff}}]{Wahlstrand07}%
  \BibitemOpen
  \bibfield  {author} {\bibinfo {author} {\bibfnamefont {J.~K.}\ \bibnamefont
  {Wahlstrand}}, \bibinfo {author} {\bibfnamefont {J.~T.}\ \bibnamefont
  {Willits}}, \bibinfo {author} {\bibfnamefont {T.~R.}\ \bibnamefont
  {Schibli}}, \bibinfo {author} {\bibfnamefont {C.~R.}\ \bibnamefont {Menyuk}},
  \ and\ \bibinfo {author} {\bibfnamefont {S.~T.}\ \bibnamefont {Cundiff}},\
  }\href {\doibase 10.1364/OL.32.003426} {\bibfield  {journal} {\bibinfo
  {journal} {Opt. Lett.}\ }\textbf {\bibinfo {volume} {32}},\ \bibinfo {pages}
  {3426} (\bibinfo {year} {2007})}\BibitemShut {NoStop}%
\bibitem [{\citenamefont {Meiser}\ and\ \citenamefont
  {Holland}(2010{\natexlab{b}})}]{MEH10}%
  \BibitemOpen
  \bibfield  {author} {\bibinfo {author} {\bibfnamefont {D.}~\bibnamefont
  {Meiser}}\ and\ \bibinfo {author} {\bibfnamefont {M.~J.}\ \bibnamefont
  {Holland}},\ }\href {\doibase 10.1103/PhysRevA.81.033847} {\bibfield
  {journal} {\bibinfo  {journal} {Phys. Rev. A}\ }\textbf {\bibinfo {volume}
  {81}},\ \bibinfo {pages} {033847} (\bibinfo {year}
  {2010}{\natexlab{b}})}\BibitemShut {NoStop}%
\bibitem [{\citenamefont {Siegman}(1986)}]{siegman86}%
  \BibitemOpen
  \bibfield  {author} {\bibinfo {author} {\bibfnamefont {A.~E.}\ \bibnamefont
  {Siegman}},\ }\href@noop {} {\emph {\bibinfo {title} {Lasers}}},\ \bibinfo
  {edition} {1st}\ ed.\ (\bibinfo  {publisher} {University Science Books, Sausalito, CA},\
  \bibinfo {year} {1986})\BibitemShut {NoStop}%
\bibitem [{\citenamefont {Lambrecht}\ \emph {et~al.}(1995)\citenamefont
  {Lambrecht}, \citenamefont {Courty}, \citenamefont {Reynaud},\ and\
  \citenamefont {Giacobino}}]{LCRG95}%
  \BibitemOpen
  \bibfield  {author} {\bibinfo {author} {\bibfnamefont {A.}~\bibnamefont
  {Lambrecht}}, \bibinfo {author} {\bibfnamefont {J.~M.}\ \bibnamefont
  {Courty}}, \bibinfo {author} {\bibfnamefont {S.}~\bibnamefont {Reynaud}}, \
  and\ \bibinfo {author} {\bibfnamefont {E.}~\bibnamefont {Giacobino}},\ }\href
  {\doibase 10.1007/BF01135854} {\bibfield  {journal} {\bibinfo  {journal}
  {Appl. Phys. B}\ }\textbf {\bibinfo {volume} {60}},\ \bibinfo {pages} {129}
  (\bibinfo {year} {1995})}\BibitemShut {NoStop}%
\end{thebibliography}

\begin{thebibliography}{7}%
\makeatletter
\providecommand \@ifxundefined [1]{%
 \@ifx{#1\undefined}
}%
\providecommand \@ifnum [1]{%
 \ifnum #1\expandafter \@firstoftwo
 \else \expandafter \@secondoftwo
 \fi
}%
\providecommand \@ifx [1]{%
 \ifx #1\expandafter \@firstoftwo
 \else \expandafter \@secondoftwo
 \fi
}%
\providecommand \natexlab [1]{#1}%
\providecommand \enquote  [1]{``#1''}%
\providecommand \bibnamefont  [1]{#1}%
\providecommand \bibfnamefont [1]{#1}%
\providecommand \citenamefont [1]{#1}%
\providecommand \href@noop [0]{\@secondoftwo}%
\providecommand \href [0]{\begingroup \@sanitize@url \@href}%
\providecommand \@href[1]{\@@startlink{#1}\@@href}%
\providecommand \@@href[1]{\endgroup#1\@@endlink}%
\providecommand \@sanitize@url [0]{\catcode `\\12\catcode `\$12\catcode
  `\&12\catcode `\#12\catcode `\^12\catcode `\_12\catcode `\%12\relax}%
\providecommand \@@startlink[1]{}%
\providecommand \@@endlink[0]{}%
\providecommand \url  [0]{\begingroup\@sanitize@url \@url }%
\providecommand \@url [1]{\endgroup\@href {#1}{\urlprefix }}%
\providecommand \urlprefix  [0]{URL }%
\providecommand \Eprint [0]{\href }%
\providecommand \doibase [0]{http://dx.doi.org/}%
\providecommand \selectlanguage [0]{\@gobble}%
\providecommand \bibinfo  [0]{\@secondoftwo}%
\providecommand \bibfield  [0]{\@secondoftwo}%
\providecommand \translation [1]{[#1]}%
\providecommand \BibitemOpen [0]{}%
\providecommand \bibitemStop [0]{}%
\providecommand \bibitemNoStop [0]{.\EOS\space}%
\providecommand \EOS [0]{\spacefactor3000\relax}%
\providecommand \BibitemShut  [1]{\csname bibitem#1\endcsname}%
\let\auto@bib@innerbib\@empty
\bibitem [{\citenamefont {Bohnet}\ \emph {et~al.}(2012)\citenamefont {Bohnet},
  \citenamefont {Chen}, \citenamefont {Weiner}, \citenamefont {Meiser},
  \citenamefont {Holland},\ and\ \citenamefont {Thompson}}]{BCW12}%
  \BibitemOpen
  \bibfield  {author} {\bibinfo {author} {\bibfnamefont {J.~G.}\ \bibnamefont
  {Bohnet}}, \bibinfo {author} {\bibfnamefont {Z.}~\bibnamefont {Chen}},
  \bibinfo {author} {\bibfnamefont {J.~M.}\ \bibnamefont {Weiner}}, \bibinfo
  {author} {\bibfnamefont {D.}~\bibnamefont {Meiser}}, \bibinfo {author}
  {\bibfnamefont {M.~J.}\ \bibnamefont {Holland}}, \ and\ \bibinfo {author}
  {\bibfnamefont {J.~K.}\ \bibnamefont {Thompson}},\ }\href@noop {} {\bibfield
  {journal} {\bibinfo  {journal} {Nature}\ }\textbf {\bibinfo {volume} {484}},\
  \bibinfo {pages} {78} (\bibinfo {year} {2012})}\BibitemShut {NoStop}%
\bibitem [{\citenamefont {Chen}\ \emph {et~al.}(2011)\citenamefont {Chen},
  \citenamefont {Bohnet}, \citenamefont {Sankar}, \citenamefont {Dai},\ and\
  \citenamefont {Thompson}}]{CBS11}%
  \BibitemOpen
  \bibfield  {author} {\bibinfo {author} {\bibfnamefont {Z.}~\bibnamefont
  {Chen}}, \bibinfo {author} {\bibfnamefont {J.~G.}\ \bibnamefont {Bohnet}},
  \bibinfo {author} {\bibfnamefont {S.~R.}\ \bibnamefont {Sankar}}, \bibinfo
  {author} {\bibfnamefont {J.}~\bibnamefont {Dai}}, \ and\ \bibinfo {author}
  {\bibfnamefont {J.~K.}\ \bibnamefont {Thompson}},\ }\href {\doibase
  10.1103/PhysRevLett.106.133601} {\bibfield  {journal} {\bibinfo  {journal}
  {Phys. Rev. Lett.}\ }\textbf {\bibinfo {volume} {106}},\ \bibinfo {pages}
  {133601} (\bibinfo {year} {2011})}\BibitemShut {NoStop}%
\bibitem [{\citenamefont {Tanji-Suzuki}\ \emph {et~al.}(2011)\citenamefont
  {Tanji-Suzuki}, \citenamefont {Leroux}, \citenamefont {Schleier-Smith},
  \citenamefont {Cetina}, \citenamefont {Grier}, \citenamefont {Simon},\ and\
  \citenamefont {Vuleti\ifmmode~\acute{c}\else
  \'{c}\fi{}}}]{TanjiSuzuki2011201}%
  \BibitemOpen
  \bibfield  {author} {\bibinfo {author} {\bibfnamefont {H.}~\bibnamefont
  {Tanji-Suzuki}}, \bibinfo {author} {\bibfnamefont {I.~D.}\ \bibnamefont
  {Leroux}}, \bibinfo {author} {\bibfnamefont {M.~H.}\ \bibnamefont
  {Schleier-Smith}}, \bibinfo {author} {\bibfnamefont {M.}~\bibnamefont
  {Cetina}}, \bibinfo {author} {\bibfnamefont {A.~T.}\ \bibnamefont {Grier}},
  \bibinfo {author} {\bibfnamefont {J.}~\bibnamefont {Simon}}, \ and\ \bibinfo
  {author} {\bibfnamefont {V.}~\bibnamefont {Vuleti\ifmmode~\acute{c}\else
  \'{c}\fi{}}},\ }\href {\doibase 10.1016/B978-0-12-385508-4.00004-8}
  {\bibfield  {journal} {\bibinfo  {journal} {Adv. At. Mol. Opt. Phys.}\
  }\textbf {\bibinfo {volume} {60}},\ \bibinfo {pages} {201} (\bibinfo {year}
  {2011})}\BibitemShut {NoStop}%
\bibitem [{\citenamefont {Meiser}\ \emph {et~al.}(2009)\citenamefont {Meiser},
  \citenamefont {Ye}, \citenamefont {Carlson},\ and\ \citenamefont
  {Holland}}]{MYC09}%
  \BibitemOpen
  \bibfield  {author} {\bibinfo {author} {\bibfnamefont {D.}~\bibnamefont
  {Meiser}}, \bibinfo {author} {\bibfnamefont {J.}~\bibnamefont {Ye}}, \bibinfo
  {author} {\bibfnamefont {D.~R.}\ \bibnamefont {Carlson}}, \ and\ \bibinfo
  {author} {\bibfnamefont {M.~J.}\ \bibnamefont {Holland}},\ }\href {\doibase
  10.1103/PhysRevLett.102.163601} {\bibfield  {journal} {\bibinfo  {journal}
  {Phys. Rev. Lett.}\ }\textbf {\bibinfo {volume} {102}},\ \bibinfo {pages}
  {163601} (\bibinfo {year} {2009})}\BibitemShut {NoStop}%
\bibitem [{\citenamefont {Brion}\ \emph {et~al.}(2007)\citenamefont {Brion},
  \citenamefont {Pedersen},\ and\ \citenamefont {M\o{}lmer}}]{BPM07}%
  \BibitemOpen
  \bibfield  {author} {\bibinfo {author} {\bibfnamefont {E.}~\bibnamefont
  {Brion}}, \bibinfo {author} {\bibfnamefont {L.~H.}\ \bibnamefont {Pedersen}},
  \ and\ \bibinfo {author} {\bibfnamefont {K.}~\bibnamefont {M\o{}lmer}},\
  }\href@noop {} {\bibfield  {journal} {\bibinfo  {journal} {J. Phys. A}\
  }\textbf {\bibinfo {volume} {40}},\ \bibinfo {pages} {1033} (\bibinfo {year}
  {2007})}\BibitemShut {NoStop}%
\bibitem [{mod()}]{modelpaper}%
  \BibitemOpen
  \href@noop {} {\enquote {\bibinfo {title} {Manuscript in preparation},}\
  }\BibitemShut {NoStop}%
\bibitem [{\citenamefont {Bohnet}\ \emph {et~al.}()\citenamefont {Bohnet},
  \citenamefont {Chen}, \citenamefont {Weiner}, \citenamefont {Cox},\ and\
  \citenamefont {Thompson}}]{BCW12Hybrid}%
  \BibitemOpen
  \bibfield  {author} {\bibinfo {author} {\bibfnamefont {J.~G.}\ \bibnamefont
  {Bohnet}}, \bibinfo {author} {\bibfnamefont {Z.}~\bibnamefont {Chen}},
  \bibinfo {author} {\bibfnamefont {J.~W.}\ \bibnamefont {Weiner}}, \bibinfo
  {author} {\bibfnamefont {K.~C.}\ \bibnamefont {Cox}}, \ and\ \bibinfo
  {author} {\bibfnamefont {J.~K.}\ \bibnamefont {Thompson}},\ }\href@noop {} {\
  }\Eprint {http://arxiv.org/abs/1208.1710v2} {arXiv:1208.1710v2
  [physics.atom-ph]} \BibitemShut {NoStop}%
\end{thebibliography}
\end{document}